\documentclass[12pt]{article}
\usepackage{epsfig}

\newcommand{\cF}{{\cal F}}
\newcommand{\pt}{\widetilde{p}}

\newcommand{\dE}{\delta {\cal E}}

\newcommand{\EQ}{\begin{equation}}
\newcommand{\EEQ}{\end{equation}}
\newcommand{\EA}{\begin{eqnarray}}
\newcommand{\EEA}{\end{eqnarray}}

\begin{document}

~
\vfill
\begin{center}
{ \Large\bf Scaling Laws and Similarity Detection}\\
\vspace{10pt}
 {\Large \bf in Sequence Alignment with Gaps}
\\
\vspace{30pt}
\renewcommand{\thefootnote}{\fnsymbol{footnote}}
Dirk Drasdo$^{(1)}$,
Terence Hwa$^{(2)}$, and
Michael L\"assig$^{(1)}$\footnote[2]{
{\em Corresponding author.} Email: lassig@mpikg-teltow.mpg.de.
Fax: +49 3328 46215.}

\renewcommand{\thefootnote}{\arabic{footnote}}

\vspace{24pt}

{$^{(1)}$ Max-Planck Institut f\"ur Kolloid- und\\
Grenzfl\"achenforschung, \\
Kantstr. 55, 14513 Teltow, Germany \\

\vspace{12pt}

$^{(2)}$ Department of Physics\\
University of California at San Diego\\
La Jolla, CA 92093-0319}
\end{center}

\vspace{1cm}

\begin{abstract}
We study the problem of  similarity detection
by  sequence alignment with gaps, using a recently established
theoretical framework based on the morphology of alignment
paths. Alignments of sequences without mutual correlations are
found to have scale-invariant statistics. This is the basis for
a scaling theory of alignments of correlated sequences. Using a simple
Markov model of evolution, we generate sequences with well-defined mutual
correlations and  quantify the {\em fidelity} of an alignment in an
unambiguous way. The scaling theory predicts the dependence
of the fidelity on the alignment parameters and on the statistical
evolution parameters characterizing the sequence correlations.
Specific criteria for the optimal choice of alignment parameters
emerge from this theory. The
results are verified by extensive numerical simulations.

\vspace{0.5cm}


\vspace{0.5cm}
\noindent 
Related (p)reprints available at
{\it http://matisse.ucsd.edu/$\sim$hwa/pub.html}.

\end{abstract}

\vfill

\newpage
\pagestyle{plain} \pagenumbering{arabic}

\section{Introduction}
Sequence alignment has been one of the most valuable computational tools
in molecular biology. It has been
used extensively in discovering and understanding functional and
evolutionary relationships among genes and proteins.
There are
two basic types of alignment algorithms: algorithms without gaps such as
BLAST and FASTA (Altschul {\it et al.}, 1990), and algorithms with gaps, for
example, the
Smith-Waterman local alignment algorithm (Smith and Waterman, 1981).
Gapless alignment is widely used in database searches because the
algorithms are fast (Altschul {\it et al.}, 1990) (computational time scales
linearly
with sequence length), the results depend only weakly on the choice of
scoring systems (Altschul, 1993), and the statistical significance of the
results is well-characterized (Arratia, {\it et al.}, 1988;
Karlin and Altschul, 1990; Karlin and Altschul, 1993).
However, gapless alignment is not sensitive to weak sequence similarities
(Pearson, 1991).
For a detailed analysis, algorithms with gaps are therefore
needed (Waterman, 1989; 1994).

At present, there are two main obstacles to the wider application
of these more powerful tools. They require substantially longer computational
time
than gapless alignments (depending quadratically on the sequence length).
As computational power continues its exponential growth at a rate even faster
than the growth of genomic information, we expect this constraint to
become less stringent in the near future. More importantly, gapped
alignments lack a detailed statistical theory assessing the significance
of the results. It is this second problem we address in the present
paper.

The common algorithms assign a score to
each alignment of two or more sequences. The score is based on the
number of matches, mismatches, and gaps. Maximization of this score is then
used to select the optimal alignment, taken as a measure of the mutual
correlations between the sequences. However, it is well known that
the optimal alignment of a given pair of sequences strongly depends on
the scoring parameters used. The same is true for the {\em fidelity}
of the optimal alignment, that is,
the extent to which mutual correlations are recovered.
Hence, the key problem of alignment statistics is to quantify the
degree of sequence similarity based on alignment data  and
to find the scoring  parameters producing alignments of the highest
fidelity. This problem has been addressed for gapless
alignments (Altschul, 1993), based on the knowledge of the exact probability
distribution function for the optimal scores in  gapless alignments
of mutually uncorrelated sequences (Karlin and  Altschul, 1990).
For algorithms with gaps, however, not even the leading moments
of the distribution function have been known so far.
Scoring parameters  have been
chosen mostly by trial and error, although  there have
been systematic efforts to establish a more solid empirical
footing (Benner, 1993; Vingron and  Waterman, 1994; Koretke {\it et al.},
1996).

To guide the choice of scoring parameters, a quantitative measure of the
fidelity of an alignment is necessary. Since the algorithm is designed to
detect residual similarities between sequences in a divergent evolution, it is
clear that the fidelity measure has to emerge from the underlying evolution
process. We use a simple probabilistic evolution model to generate daughter
sequences from  ancestor sequences by local substitutions, insertions, and
deletions.  The model is certainly too simple to describe realistic evolution
processes, but it allows an unambiguous identification of inherited mutual
similarities between  sequences. The fidelity of an alignment is then simply
the fraction of the inherited similarities recovered by it. Maximization of the
fidelity is used as a criterion to select optimal scoring parameters.

We will not address here algorithmic and computational
aspects of alignments. Efficient algorithms are available for parametric
and ensemble alignment (Waterman {\it et al.}, 1992,
Gusfield {\it et al.}, 1992, Waterman, 1994).
Our goal is to present a {\em statistical theory}  of gapped
alignments. This theory can then be used to {\em predict } optimal
scoring parameters appropriate for different classes
of inter-sequence correlations.

As is well recognized, the main mathematical difficulty preventing a
quantitative
statistical characterization  of gapped local alignment
lies within the {\em global} alignment regime. In two recent
communications
(Hwa and  L\"assig, 1998; Drasdo {\it et al.} 1998), we have shown that
the statistical properties in the parameter regime
close to the log-linear phase transition line (Waterman {\it et al.}, 1987;
Arratia and  Waterman, 1994)
of the Smith-Waterman local alignment algorithm are in fact dominated by
the statistics of global alignment. This regime is important for
biological applications since it has been found empirically to
produce ``good'' alignments (Vingron and
Waterman, 1994). It is thus very important to
characterize the statistics of global alignment, which is the purpose of
this paper. We report a detailed study of the  properties
of global alignments of mutually {\em uncorrelated} as well as {\em correlated}
sequences by the Needleman-Wunsch (1970) algorithm. The results
are used to select optimal scoring parameter to detect sequence correlations
generated by the toy evolution process. They can also be incorporated
directly into the parameter selection procedure for local alignment
(Hwa and  L\"assig (1998); Drasdo {\it et al.}, 1998).

The statistical theory of gapped alignments presented here is based on
a {\em geometrical} approach introduced recently by two of us
(Hwa and  L\"{a}ssig, 1996).
This   approach focuses on the morphology of the optimal
{\em alignment paths}.
The notion of an alignment path (recalled below)
provides a very fruitful link to various well-studied problems of
statistical mechanics (Kardar, 1987; Fisher and  Huse, 1991;  Hwa and  Fisher,
1994)
as has also been noticed by Zhang and Marr (1995).
The important statistical properties of alignment paths are
described by a number of {\em scaling laws} (Hwa and L\"{a}ssig, 1996;
Drasdo {\it et al.}, 1997)
explained in detail below. Their applicability
to  alignment algorithms is supported by extensive
numerical evidence. The resulting scaling theory of alignment has three
main virtues: \newline
(i) It distinguishes clearly between {\em universal}
(parameter-independent) properties of alignments and those depending
on the scoring parameters (and hence governing their optimal choice). \newline
(ii) It relates score data of alignments to their fidelity
and to the underlying evolutionary parameters characterizing the
similarities of the sequences compared. \newline
(iii) Its key statistical {\em averages} turn out to be significant
for the alignment of {\em single} sequence pairs that are sufficiently
long. \newline
Statistical scaling theories have also been developed for related
optimization problems in structural biology, notably
protein folding (Wang {\it et al.}, 1996; Onuchic {\it et al.}, 1997).

This paper is organized as follows. In Section~2, we define the evolution
process, recall the global alignment algorithm used throughout this paper,
and discuss the qualitative aspects of the geometrical approach. The
quantitative theory of alignment starts in Section~3, where
we give a detailed description of the alignment statistics for uncorrelated
random sequences, and present the power laws governing
alignment paths and scores. In Section~4, we turn to sequences with mutual
correlations inherited by a realization of
our evolution process. We establish a scaling theory
that explains the
parameter dependence of alignments in a quantitative way. Hence we
derive optimal alignment parameters as a function of the evolution
parameters, i.e., the frequency of indels and
substitutions\footnote{A conceptually similar link between
scoring parameters and evolution parameters has been discussed in the
context of maximum-likelihood methods (Bishop and  Thompson, 1986;
Thorne {\it et al.}, 1991, 1992).}.
Furthermore, we
show how the evolutionary parameters and the optimal alignment
of a given pair of sequences can be deduced from its score data.


\section{The geometrical approach to sequence alignment}


\subsubsection*{Evolution model}

The  evolution process used in this paper
evolves from an ``ancestor'' sequence $Q$
of length $N \gg 1$
whose elements are labeled by the index $i$. The element $Q_i$
is chosen from a set of $c$ different letters.  Each letter occurs
with equal probability $1/c$, independently of the elements
at other positions.
Hence, the ancestor sequence is a Markov random sequence.
The numerical results presented below are for the case
$c = 4$ as appropriate for nucleotide sequences, but for some
derivations, it is useful to consider general $c$-letter alphabets.

The evolution process generates a daughter sequence $Q'$ of length $N'$
from the ancestor sequence $Q$. It involves local insertions and deletions
of random elements with the same probability $\pt$, and point substitutions
by a random element with probability $p$. Insertion, deletion, and
substitution events at one point of the sequence are independent of the events
at other points. The evolution process can thus be formulated as a
Markov process along the sequence
(Bishop and  Thompson, 1986, Thorne {\it et al.}, 1991; Hwa and  L\"assig,
1996).
The precise evolution rules used in this paper are given in Appendix A.
These rules are such that the average length of the daughter sequence,
$\overline N$, equals the length $N$ of the ancestor sequence.

A specific realization of this Markov process defines a unique
{\em evolution path} linking the sequences $Q$ and $Q'$; see Fig.~1(a).
However, the same pair of sequences can
be linked by different evolution paths.
Any evolution path has a number of conserved elements (i.e., elements
that are not deleted or substituted at any point of the evolution
process). The average fraction of ancestor elements $Q_i$ conserved in
the daughter sequence $Q'$ is
\begin{equation}
U(p,q) = (1 - p)\,(1 - q) \;,
\label{U}
\end{equation}
where
\begin{equation}
q = \frac{\pt}{1 - \pt}
\label{q}
\end{equation}
is the effective insertion/deletion rate
(see Appendix~A).
We call these conserved pairs of elements $(Q_i = Q'_j)$
{\em native pairs}. Their fraction $U(p,q)$ quantifies the mutual similarity
between sequences. In the remainder of this paper, we take
$U$ and $q$ as the basic parameters characterizing the evolution
process. The primary goals of sequence alignment
are to identify the native pairs and to estimate the mutual similarity
$U$.

\begin{figure}[t]
\begin{center}
\epsfig{file=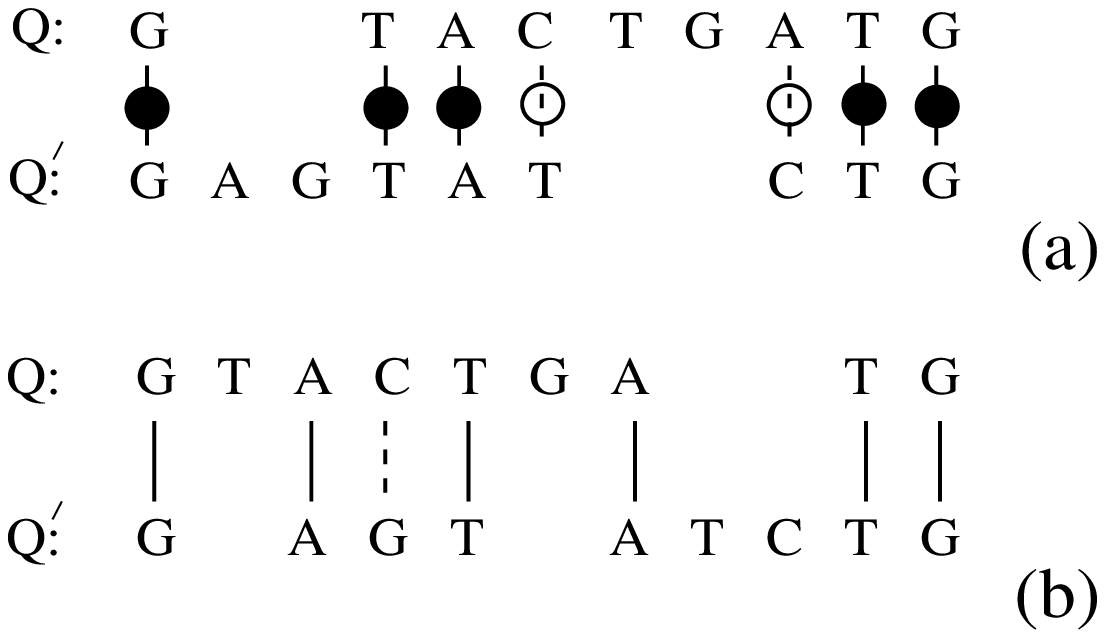,height=2in,angle=0}\hspace{0.5in}
\epsfig{file=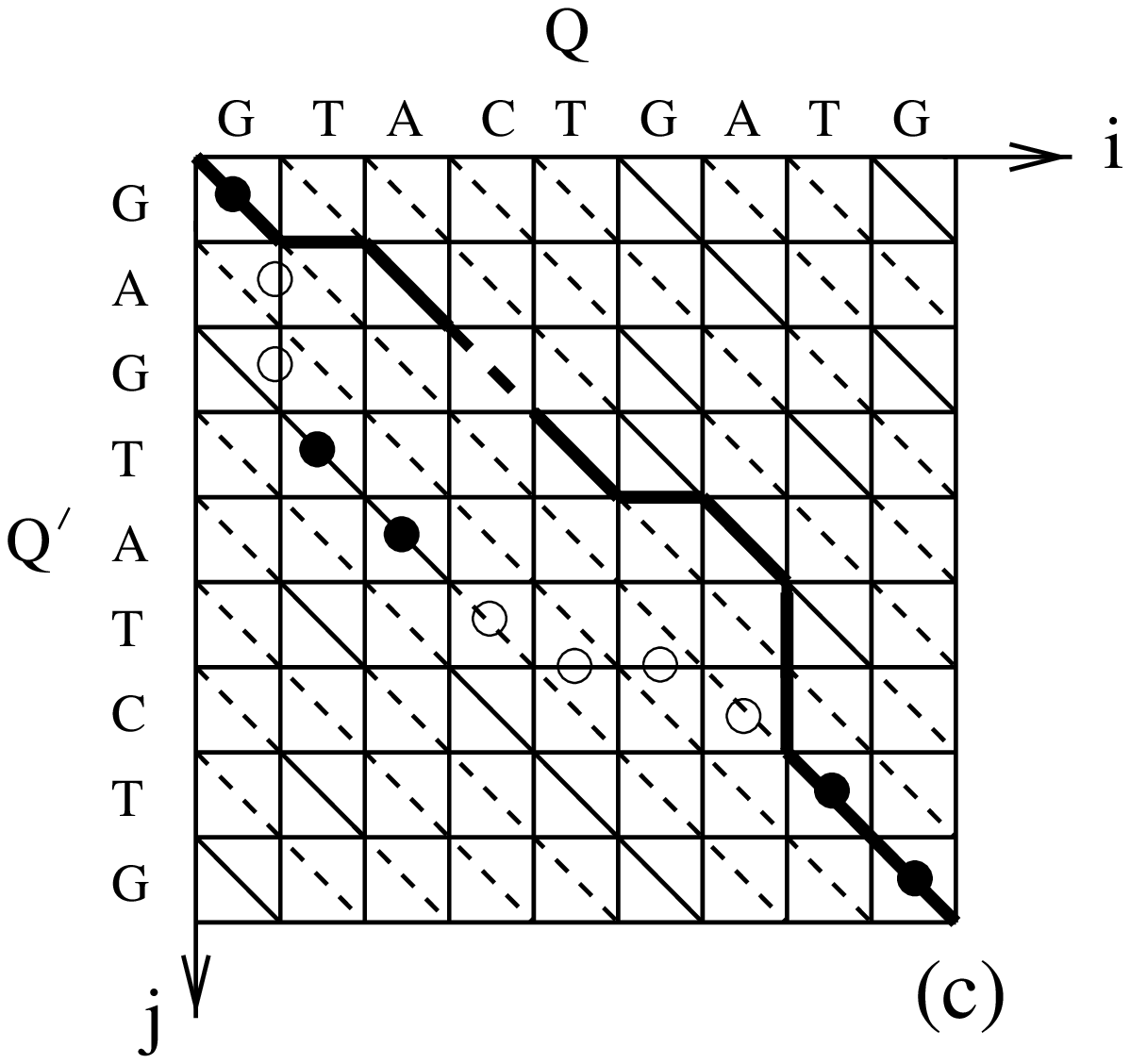,height=2 in,angle=0}
\end{center}
\vspace{12pt}
\baselineskip=12pt {\small Fig.~1:
(a) A Markov evolution path linking the  two  sequences
$Q = \{G,T,A,C,T,G,A,T,G\}$ and
$Q'= \{G,A,G,T,A,T,C,T,G\}$. Native pairs are marked by bonds with
full circles, substitutions by bonds with empty circles. The unpaired letters
$Q_i$ are deleted, the unpaired letters $Q'_j$ are inserted.
(b) A possible alignment between $Q$ and $Q'$ with matches $(Q_i = Q'_j)$
(full lines), mismatches $(Q_i \neq Q_j)$ (dashed lines) and gaps (unpaired
letters).
(c) Lattice representation. The evolution path $R(t)$ corresponding to (a)
is marked by circles; there are five native bonds (full circles).
The alignment path corresponding to (b) appears as thick line whose solid
(dashed) diagonal bonds are matches (mismatches) and whose horizontal and
vertical bonds are gaps. It covers three of the five native bonds, producing
the fidelity $\cF = 3/5$.
}
\end{figure}


\subsubsection*{Alignment and Scoring Scheme}

We align the sequences $Q = \{ Q_i \}$ and $Q' = \{Q'_j\}$
using the simplest version of the
global alignment algorithm by Needleman and Wunsch (1970).
A global alignment of two sequences is defined
as an ordered set of pairings $(Q_i, Q'_j)$ and of gaps
$(Q_i,-)$ and $(-,Q'_j)$, each element $Q_i$ and
$Q'_j$ belonging to
exactly one pairing or gap (see Fig.~1(b)). Any alignment is assigned a
score $S$, maximization of which defines the optimal alignment\footnote{
In statistical mechanics, one may think of $-S$ as an energy that has to be
minimized.}.
We use here the simplest class of {\em linear} scoring functions
(Smith and  Waterman, 1981),
with the score given by the total number $N_{+}$ of matches
$(Q_i = Q'_j)$, the total number $N_{-}$ of mismatches
$(Q_i \neq Q'_j)$, and
the total number $N_g$ of gaps. Hence, the most general such function
involves three scoring parameters:
\begin{equation}
S = \mu_+ N_+ + \mu_- N_- + \mu_g N_g \;.
\label{Sgen}
\end{equation}
However, the {\em optimal} alignment configuration
 of a given sequence pair $Q$ and $Q'$ is left invariant if the three
 scoring parameters are all multiplied
 by the same factor. Along with the property that $2N_+ + 2N_- + N_g = N$
is conserved in global alignment and the invariance of the alignment
configuation to an additive constant to (\ref{Sgen}), we see that
the outcome of global alignment is controlled effectively by a {\em single}
parameter..   Without loss of generality\footnote{
Indeed, the scoring functions (\ref{Sgen}) and (\ref{S}) lead to the
same optimal alignment if
\[
\gamma = \frac{c - 2}{ 2 \sqrt{c - 1} } +
	 \frac{c}{ 2 \sqrt{c - 1} } \frac{\mu_+ + \mu_- - 2
      \mu_g}{\mu_+ - \mu_-} \;.
\]}, we may
 therefore choose to use the scoring
function
\EQ
S = \sqrt{c - 1}\, N_+ - \frac{1}{\sqrt{c - 1}} \, N_- - \gamma N_g \;,
\label{S}
\EEQ
which is normalized in such a
way that a pairing of two independent random
elements has the average score $0$ and the score variance $1$.
The scoring function $S$ depends only on the parameter $\gamma$, which
describes the effective cost of a gap over pairing.
The optimal alignment depends on $\gamma$ in the regime
$\gamma \geq \gamma _0\equiv 1/(2\sqrt{c-1})$ (i.e.,
$2 \mu_g > \mu_-$) to which we restrict ourselves in the sequel.
For $\gamma < \gamma_0$, it is always favorable to replace a mismatch
by two gaps (Waterman {\it et al.}, 1987),
and the alignment is not biological relevant.


\subsubsection*{The fidelity of an alignment}

As discussed above, mutual correlations between the sequences
$Q= \{ Q_i\}$ and $Q' = \{Q_j\}$ arise from the set of native pairs
$(Q_i = Q'_j)$. The {\em fidelity} $\cF$ of an alignment can be quantified as
the
fraction of correctly matched native pairs, see Fig.~1(b).
This is an unambiguous measure of the goodness of an alignment, and
it will be used below to find optimal alignment parameters.
To evaluate $\cF$ directly, the native pairs have to be distinguished from
random matches
$(Q_i = Q'_j)$ involving mutated elements. Hence,
the fidelity defined in this
way depends not only on the sequences $Q$ and
$Q'$ but also on the evolution path linking them.
Of course, the evolution path is not known in most applications
of sequence alignment. However, the scaling theory discussed below
relates statistical properties of $\cF$ to alignment data, making it a useful
and measurable quantity.


\subsubsection*{Lattice representation}

Any alignment of two sequences $\{Q_i\}$ and $\{Q'_j\}$ is
conveniently represented on a two-dimensional
$N\times N^{\prime }$
grid as in Fig.~1(c)
(Needleman and Wunsch, 1970). The cells of this grid are labeled by the index
pair $%
(i,j)$. The diagonal bond in cell $(i,j)$ represents the pairing of the
elements $(Q_i, Q'_j)$. The horizontal bond between cells $(i,j)$
and $(i,j+1)$ represents a gap $(Q_i, -)$ located on sequence $Q'$
between the elements $Q'_j$ and $Q'_{j+1}$.
The vertical bond between cells $(i,j)$ and $(i+1,j)$ represents a gap
located on sequence $Q$ between the elements $Q_i$ and
$Q_{i+1}$. In
this way, any alignment defines a unique {\em directed path} on the grid. Using
the rotated coordinates $r\equiv j - i$ and $t\equiv i+j$, this path is
described by a single-valued function $r(t)$ measuring the displacement
of the path from the diagonal of the alignment grid.
A path associated with an optimal alignment is denoted by $r_0(t)$.
For global alignment of typical sequences, the optimal path extends over
the entire grid, i.e., it has a length of the order $N + \overline{N'} = 2N$.
The Needleman-Wunsch dynamic programming algorithm obtains
optimal alignment paths
by computing the ``score landscape'' $S_0(r,t)$ sequentially for
all lattice points, where $S_0(r,t)$ denotes the optimal score
for the set of all alignment paths ending at the point $(r,t)$.
The version of the algorithm used in this paper is detailed in
Appendix B.

In a similar way, any evolution path linking the sequences $Q$
and $Q'$ defines a directed path $R(t)$ on the alignment grid
(called evolution path as well) (Hwa and L\"{a}ssig, 1996).
On this path, horizontal and vertical bonds represent deleted and inserted
elements, respectively. A
fraction $U$ of the bonds along the evolution path are {\em
native} bonds representing the native pairs $(Q_i = Q'_j)$.
The fidelity of an alignment is then simply the fraction of native
bonds that are also part of the corresponding alignment path $r(t)$,
see Fig.~1(c).


\subsubsection*{Alignment morphology}

Alignment algorithms are designed to trace the mutual correlations
between sequences. As it becomes clear from Figs.~2, the presence of
such correlations affects both the morphology of the optimal alignment
path $r_0(t)$ and the associated score statistics. Fig.~2(a) shows the
path $r_0(t)$ for a pair of mutually uncorrelated random sequences. This
path is seen to be intrinsically rough; i.e., the displacement has large
variations. This ``wandering'' is caused by random agglomerations of
matches in different regions of the alignment grid. Fig.~2(b) shows the
corresponding score landscape $S_0 (r,t)$ for a given value of $t$.
The maximum score value occurs at the point $r_0(t)$
and is seen to be not very pronounced; near-optimal score values occur also
at  distant points such as $r_1$. The statistics of
alignment paths and scores for uncorrelated sequences are discussed
in detail in Section~3 below.

\begin{figure}
\begin{center}
{\epsfig{figure=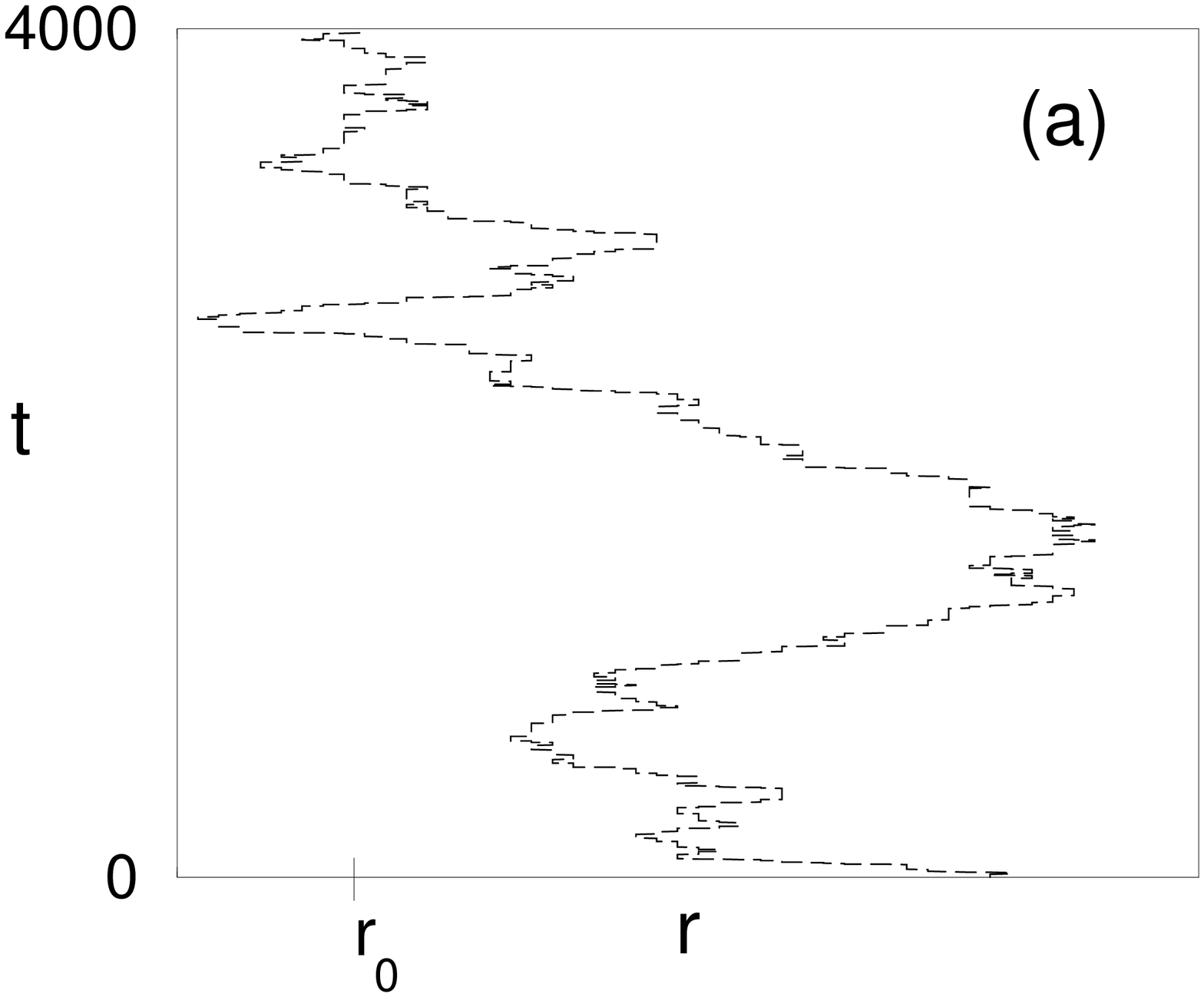,width=2.5in,angle=270}}\hfill
{\epsfig{figure=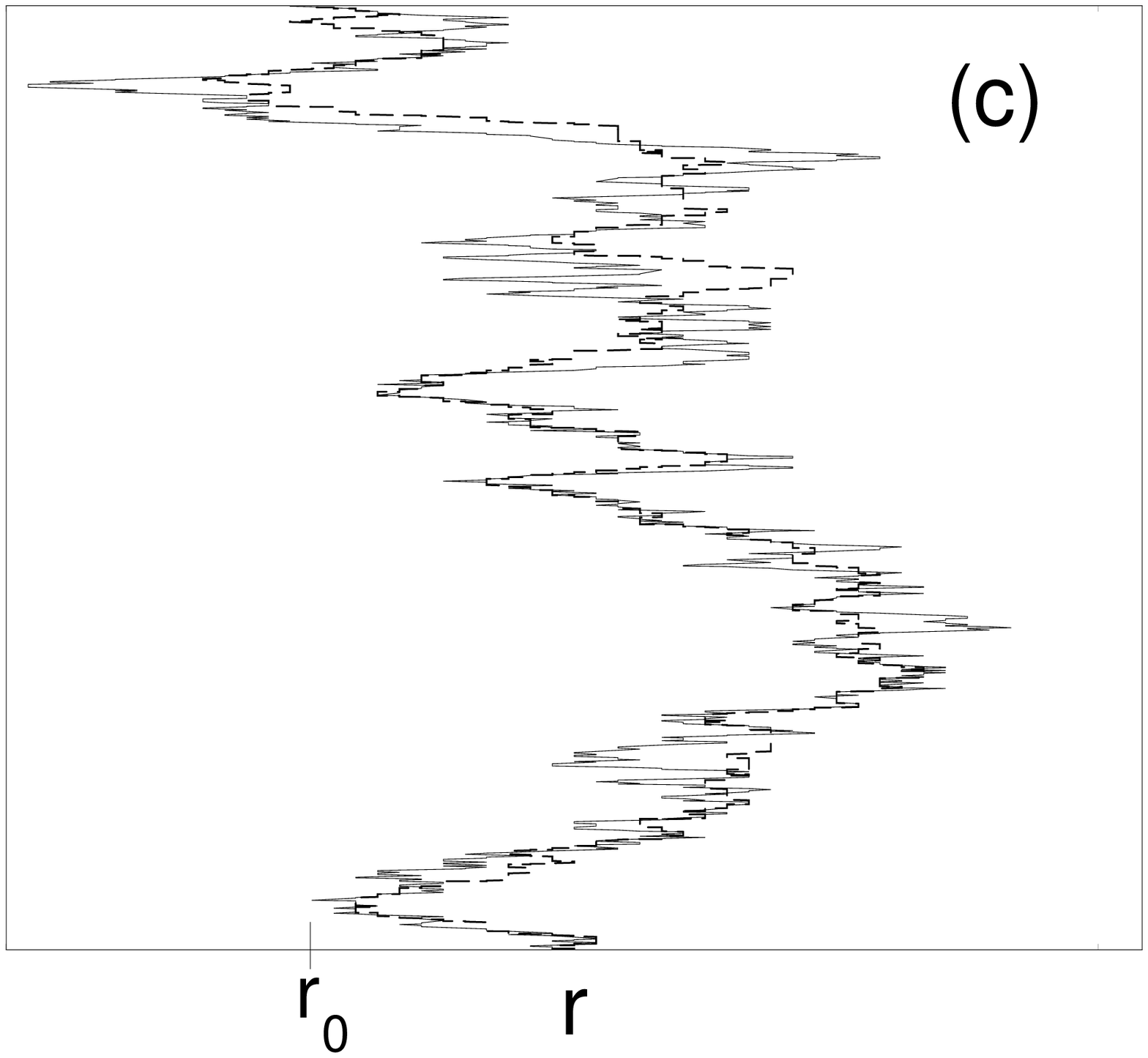,width=2.5in,angle=270}}\\
\vspace{12pt}
\epsfig{figure=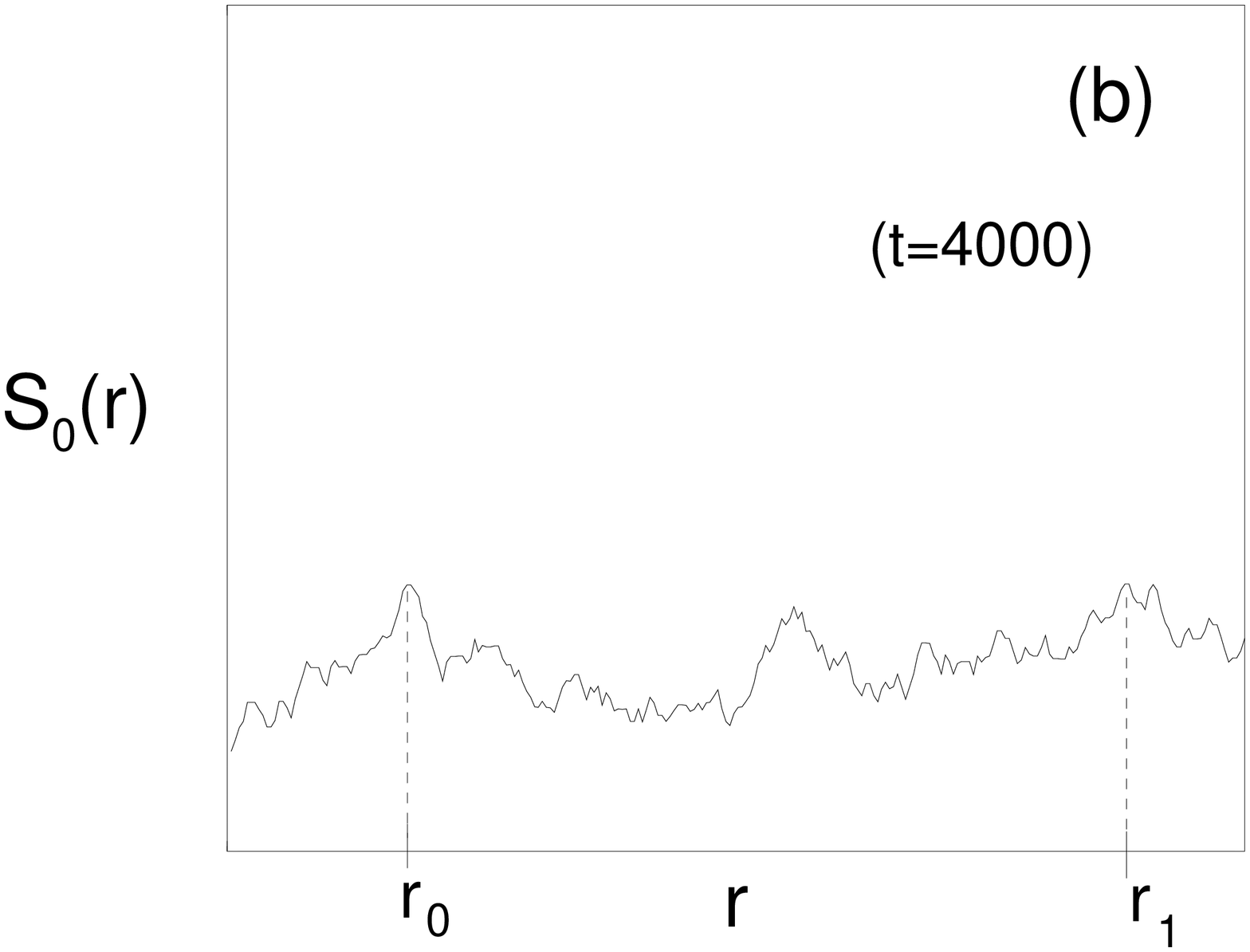,width=2.5in,angle=270}\hfill
\epsfig{figure=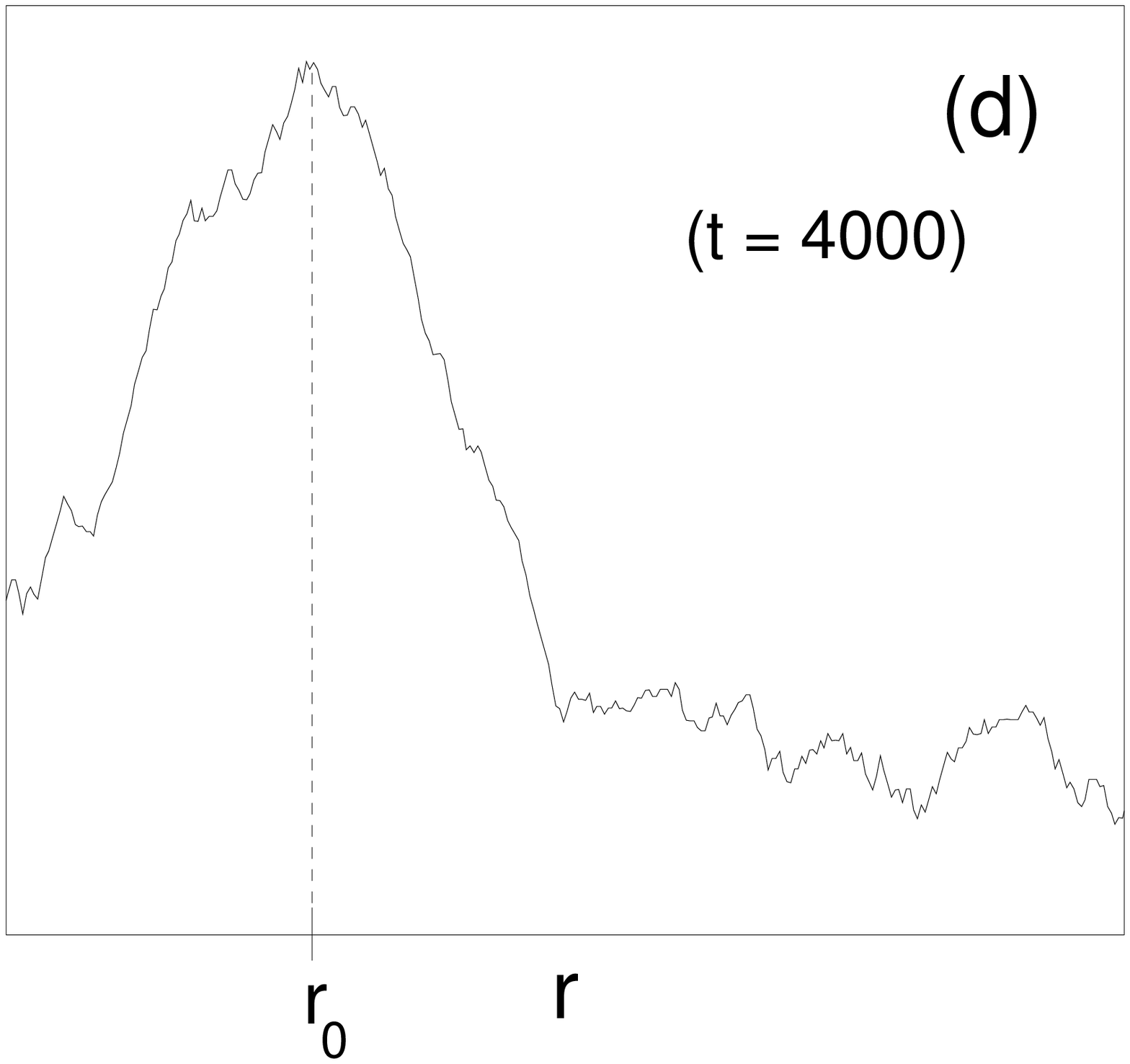,width=2.5in,angle=270}
\end{center}
\vspace{12pt}
\baselineskip=12pt {\small Fig.~2:
(a) The optimal alignment path $r_0(t)$ and (b) a slice of
the score landscape $S(r,t = 4000)$
for a pair of mutually uncorrelated random sequences.
The score maximum is at $r_0$, which
defines the endpoint $r_0 \equiv r_0(t = 4000)$
of the optimal path. Similar score values occur also at distant points
such as $r_1$.
(c) The paths $r_0(t)$ (dashed line), $R(t)$ (solid line)
and (d) the score landscape $S_0(r)$ at $t = 4000$ for
a pair of sequences with mutual correlations. The score maximum at
$r_0$ is now pronounced; all distant points $r$ have a substantially
lower score. Hence the fluctuations of the alignment path $r_0(t)$
are confined to a corridor around the evolution path $R(t)$.
}
\end{figure}

\begin{figure}
\begin{center}
\epsfig{file=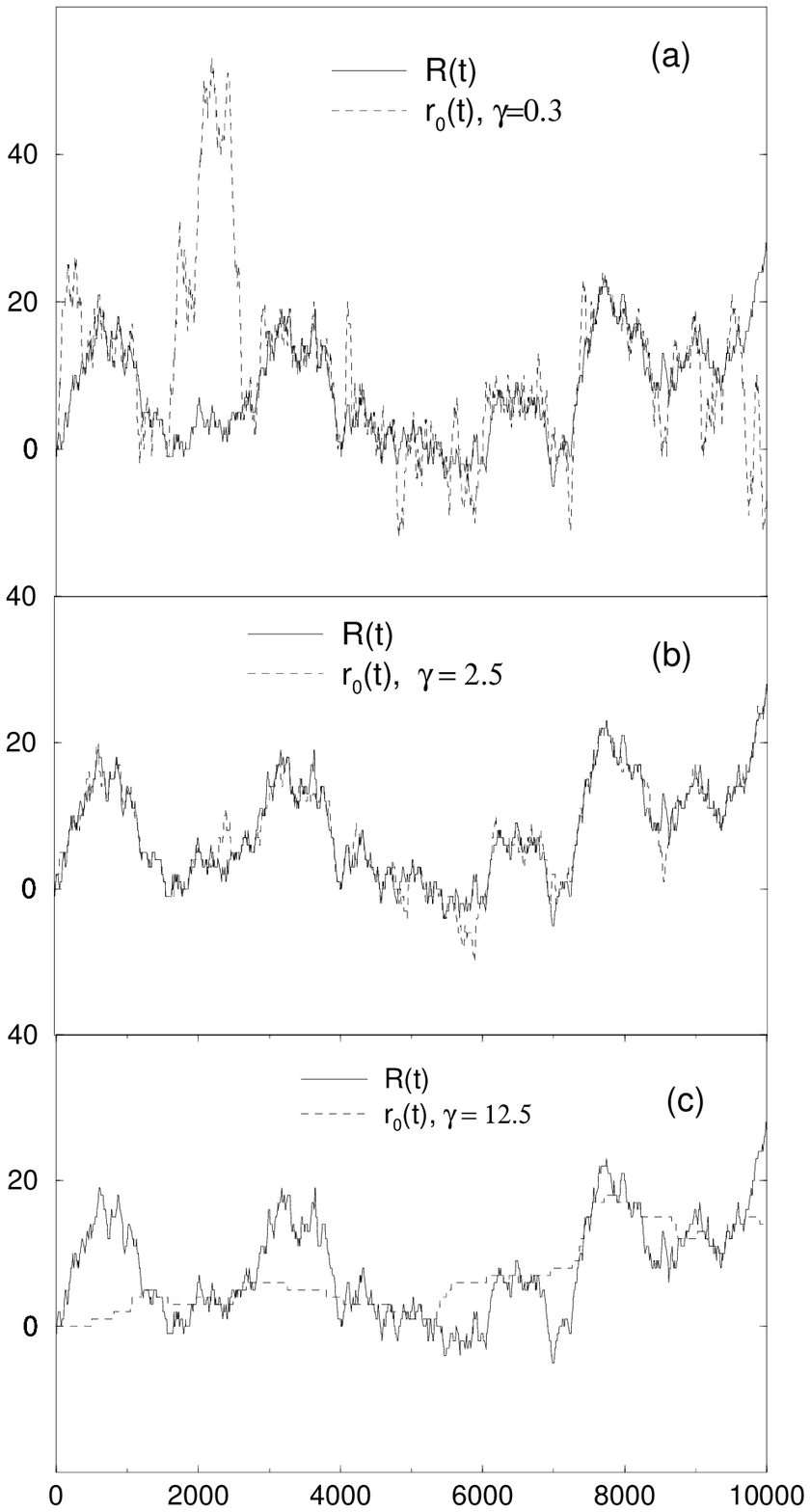,width=3in,angle=0}
\end{center}
\vspace{12pt}
\baselineskip=12pt {\small Fig.~3:
Optimal alignment paths $r_0(t)$ for the same pair of correlated
sequences and three different values of $\gamma$. The evolution
path $R(t)$ (solid lines) is the same in all three cases, while the
optimal alignment paths $r_0(t)$ (dashed lines) differ.
(a) Random fluctuation regime ($\gamma < \gamma^*)$. The
path $r_0(t)$ has strong fluctuations since the gap cost is low.
(b) Optimal alignment parameter $\gamma = \gamma^*$. The fluctuations
of the paths $r_0(t)$ and $R(t)$ are of the same order of magnitude.
(c) Shortcut regime ($\gamma > \gamma^*$). At high gap cost, the fluctuations
of $R(t)$ are dominant, while $r_0(t)$ contains large straight segments.
}
\end{figure}

The optimal alignment path for
a pair of mutually correlated sequences (obtained from the evolution
process described above) behaves quite differently, as shown in Fig.~2(c).
Its wandering is essentially restricted to a  ``corridor'' of finite
width centered around around the evolution path $R(t)$. In this way, the
path $r_0(t)$ covers a finite fraction $\cF$ of the native bonds. The
corresponding score landscape is shown in Fig.~2(d). The maximum at
$r_0(t)$ is now very pronounced;
all paths ending at points distant from $r_0(t)$
have a substantially lower score than the optimal path. The alignment
statistics of mutually correlated sequence pairs is described in
Section~4.

The morphology of the optimal alignment path depends
strongly on the choice of the scoring parameter $\gamma $. As an example,
Fig.~3 shows the optimal paths $r_0(t)$ (dashed lines) for the {\em %
same} pair of correlated  sequences with the same underlying
evolution  path $R(t)$
(the solid line), and for three different values of $\gamma $:
At small $\gamma $, the path $r_0(t)$ follows the evolution path only on large
scales. On small scales, variations in the displacement $r_0(t)$ are seen to
be larger than those of $R(t)$ (Fig.~3(a)). The intrinsic
roughness of the optimal alignment path limits its
overlap with the evolution path, hence suppressing the fidelity.
The fidelity is the highest at some intermediate value $\gamma^* $, where
the alignment path follows the target path most closely (Fig.~2(b)).
At large $\gamma $, the alignment path contains large straight segments
(Fig.~2(c)), which again reduces the fidelity.

A qualitative understanding of this parameter dependence may be gained
from an analogy to random walks, regarding $r_0(t)$ as the trajectory of
walker following a curvy path $R(t)$. The intrinsic properties
of the walker are parametrized\footnote{
In statistical mechanics, $\gamma$ is the effective line tension of
the fluctuating path $r_0(t)$.}
by $\gamma$. For small $\gamma$, the
the walker is drunk and cannot follow the path $R(t)$
without meandering to its left and right. This is the regime of
Fig.~2(a), which we call the {\em random fluctuation} regime. For large
values of $\gamma $, on the other hand, the walker is lazy and bypasses
the larger turns of the path $R(t)$; this is the {\em shortcut} regime
(Fig.~2(c)).
~From this analogy, it becomes plausible that a walker who is neither too drunk
nor too lazy will follow the path $R(t)$ most closely and thereby
achieve the highest fidelity (Fig. 2(b)). Such a criterion for the optimal
parameter $\gamma^*$ will indeed emerge from the quantitative theory
described in the remainder of this paper.


\section{Alignment of Uncorrelated Sequences}

A statistical theory of alignment can hardly predict the optimal
alignment for a specific pair of sequences. What can be characterized
are quantities averaged over realizations of the evolution process for given
parameters $U$ and $q$. It will be shown, however, that these {\em
ensemble averages} are also relevant for the alignment statistics of
single pairs of ``typical'' sequences provided they are sufficiently long.
The approach is different from the extremal statistics of the score
distribution that has been used to assess the significance of alignment
results (Karlin and  Altschul, 1990, 1993).

In the absence of mutual correlations (i.e., for $U=0$), the statistics of
alignments is determined by a balance between the loss in score due to gaps
and the gain in score due to an excess number of random matches. As
discussed by Hwa and L\"{a}ssig (1996), the corresponding alignment paths
belong to a
class of systems known in statistical mechanics as {\em directed polymers} in
a random medium~\footnote{This is also known as
the problem of first passage percolation. A detailed mathematical analysis
of the scaling laws presented below can be found in recent works by
Licea {\it et al.} (1994, 1996). The main  difference
to alignment statistics
is the number of {\em independent} random variables on a grid
of size $N \times N$. For directed polymers and first passage percolation,
this number is of order $N^2$; for
alignments of random sequences, it is only of order $N$
(see also Arratia and  Waterman, 1994).
This difference is, however, irrelevant for the asymptotic scaling
behavior (Hwa and L\"assig, unpublished). A detailed
heuristic discussion of this equivalence  in the context of a number of
closely related problems is given by Cule and Hwa (1997).}.
The statistical properties of directed polymers have been
characterized in detail (Kardar, 1987; Huse and  Fisher, 1991; Hwa and  Fisher,
1994).
We now recall the main results
and give numerical evidence of their applicability to sequence
alignment.


\subsubsection*{Displacement and score statistics}

The displacement
$\Delta r_0 (t_2 - t_1) \equiv r_0 (t_2) - r_0(t_1)$ of the optimal alignment
path between two arbitrary points $t_1$ and $t_2$
is found to obey the statistical scaling law
\begin{equation}
\overline{(\Delta r_0 (t))^2} \simeq A^2 (\gamma) \, |t|^{4/3} \;,
\label{r02}
\end{equation}
the overbar denoting the average over an ensemble of mutually
uncorrelated sequence pairs.
Eq.~(\ref{r02}) is an asymptotic law
valid for
$\overline{(\Delta r_0 (t))^2} \gg 1$, i.e., for
$ t \gg  t_0 (\gamma) \equiv A^{-3/2} (\gamma)$.
It says that the exponent $4/3$ is a
very robust feature of the optimal alignment of uncorrelated random
sequences, independent of the scoring parameter(s) or even scoring schemes
used. A large gap cost efficiently suppresses
the displacement only for the  limited range of scales $t < t_0 (\gamma)$.
On larger scales, the cost of gaps is always outweighed by the gain in score
from regions of the alignment grid with an excess number of random
matches, leading to the  power law (\ref{r02}) with a ``universal''
exponent.
The dependence of the roughness $\overline{(\Delta r_0(t))^2}$ on
the scoring parameters ($\gamma $ in this case) is contained entirely in the
amplitude $A(\gamma )$; this dependence  is discussed below.
The ensemble average (\ref{r02}) also describes the displacement {\em
auto-correlation function} of the optimal path for a {\em single} sequence
pair, defined as an average over initial points $t_1$ in an
interval $T \gg t$,
\begin{equation}
\overline{(\Delta r_0 (t))^2} =
 T^{-1} \sum_{t_1 = 1}^{T} (r_0(t_1 + t) - r_0(t_1))^2 \;.
\label{auto}
\end{equation}

The large displacement fluctuations of the optimal alignment
path $r_0(t)$ are accompanied by large variations in its score.
For an ensemble of mutually uncorrelated sequences,
the score average is asymptotically linear in the length $N$,
\begin{equation}
\overline{S_0(N,\gamma)} \simeq E_0(\gamma) \, N
\label{ave}
\end{equation}
for $N \gg 1$, with a monotonically decreasing coefficient function
$E_0(\gamma)$. However, the variance of optimal score is described
by a nontrivial power law
\begin{equation}
\overline{(\Delta S_0(N,\gamma))^2}
 \simeq B^2 (\gamma) N^{2/3}
\label{E02}
\end{equation}
which is valid in the asymptotic regime $N \gg t_0 (\gamma)$. The
dependence on the alignment parameters is again only in the amplitude
$B(\gamma)$, while the exponent $2/3$ is universal. The ensemble average
can be obtained (up to a $\gamma$-independent proportionality factor)
from a single pair of sequences as average
in the score landscape $S_0(r,t)$ over a sufficiently long interval
$r_1 \leq r \leq r_1 + R$,
\begin{equation}
\overline{(\Delta S_0(t,\gamma))^2} \sim
  R^{-1} \sum_{r = r_1}^{r_1+R-1} S^2_0(r,t) -
  \left ( R^{-1} \sum_{r = r_1}^{r_1+R-1} S_0(r,t) \right )^2 \;,
\label{Er2}
\end{equation}
see Appendix~B.
We have verified the scaling laws (\ref{r02}) and
(\ref{E02}) numerically for a range of $\gamma$ values; see Figs.~4.
The asymptotic behavior is found to set in rather quickly for
$t > t_0(\gamma)$. The same scaling has been found for a pair of
unrelated cDNA sequences (see also Fig.~4), which justifies our
modeling of individual sequences as Markov chains. A more comprehensive
study of correlated and uncorrelated cDNA sequences will be presented
elsewhere.

\begin{figure}
\begin{center}
\epsfig{figure=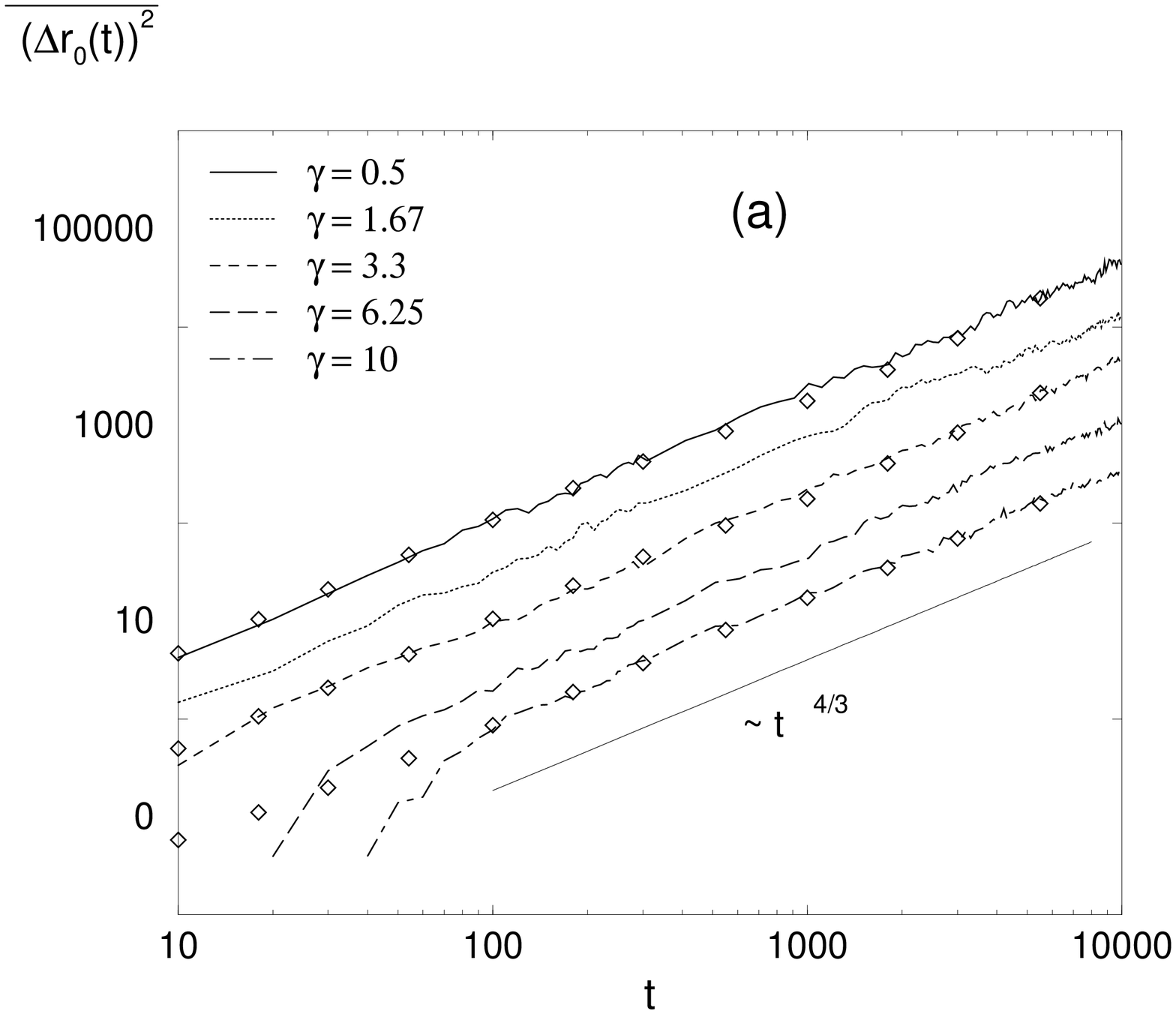,height=3.0in,angle=270}\hspace{0.3in}
\epsfig{figure=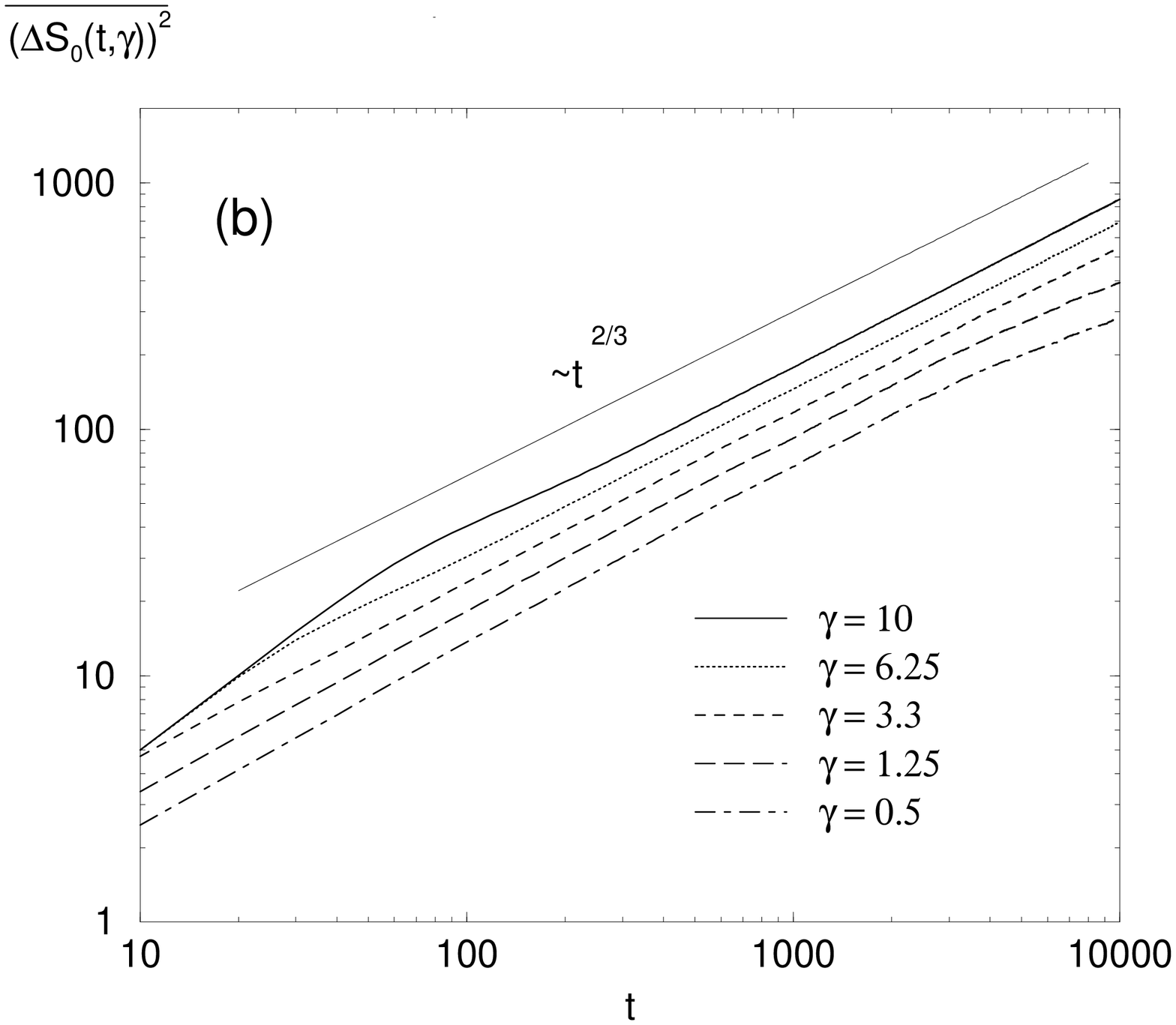,height=3.0in,angle=270}\\
\epsfig{figure=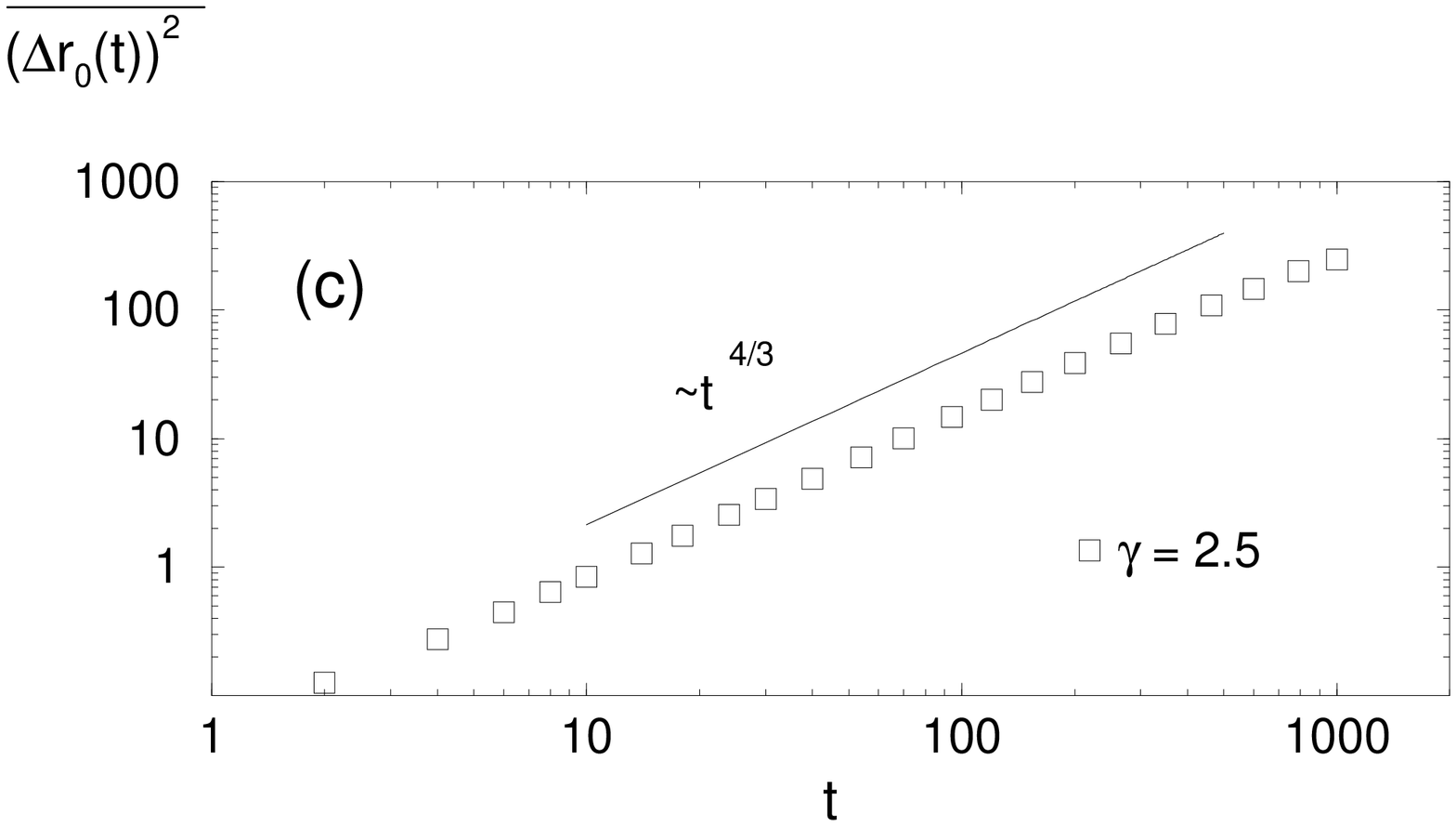,height=3.0in,angle=270}\hspace{0.3in}
\epsfig{figure=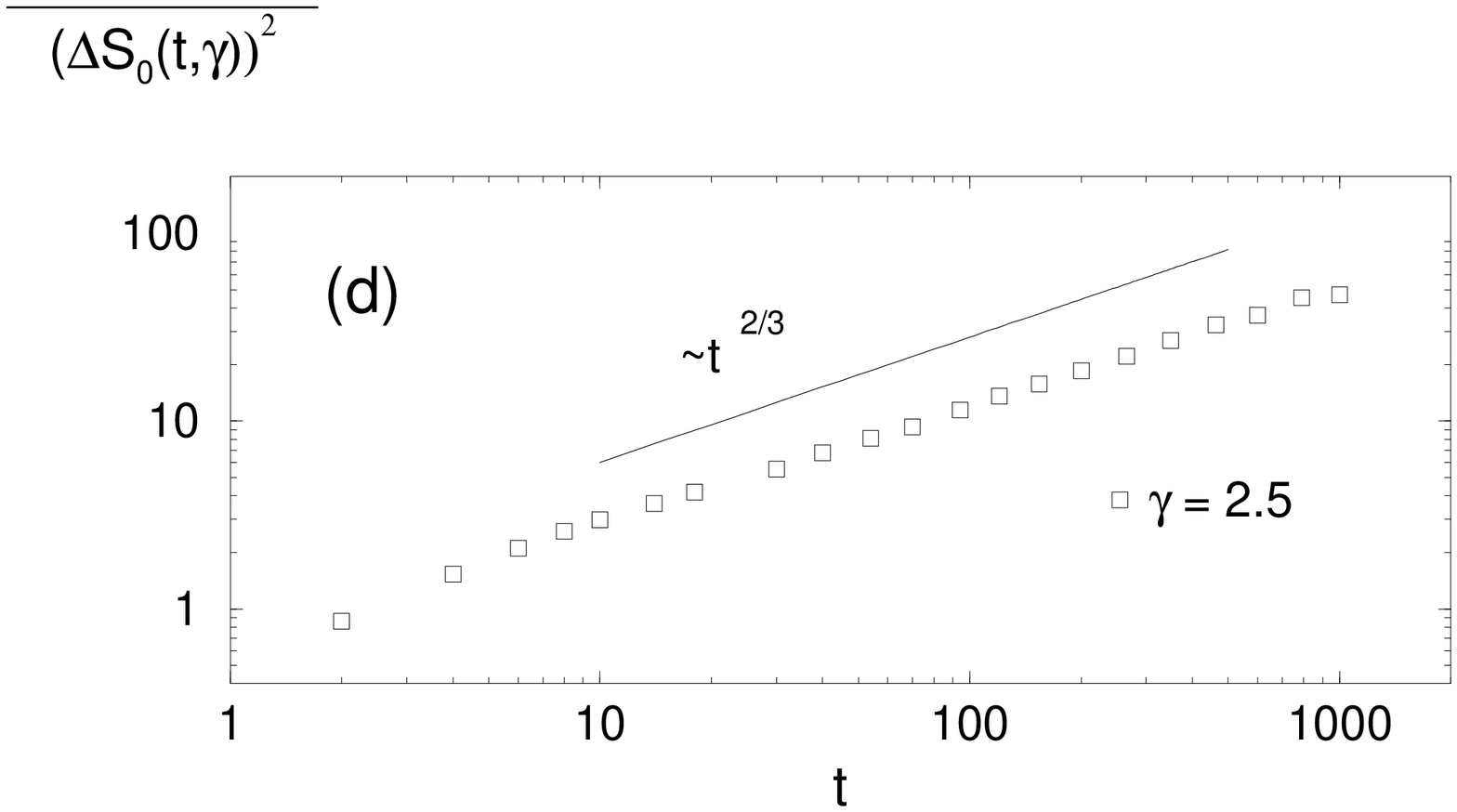,height=3.0in,angle=270}
\end{center}
\vspace{12pt}
\baselineskip=12pt {\small Fig.~4:
(a) Displacement fluctuations $\overline{(\Delta r_0 (t))^2}$ of the optimal
alignment for several values of $\gamma$.
Averages  over an ensemble of 200
mutually uncorrelated sequence pairs
are marked by lines, auto-correlation functions for a single sequence
pair of length $N = 10^5$ by squares.
(b) Score fluctuations $\overline{(\Delta S_0 (t))^2}$ obtained from
the score landscape (\ref{tm}) by Eq.~(\ref{Er2}) for several values
of $\gamma$.
(c) Displacement auto-correlation function and
(d) score fluctuations for a pair of unrelated cDNA seqences
(P.lividius cDNA for COLL2alpha gene (Exposito {\it et al.}, 1995) and
Drosophila melanogaster (cDNA1) protein 4.1 homologue (coracle)
mRNA, complete cds. (Fehon {\it et al.}, 1994)).
The straight lines indicate the expected
power laws given by Eqs.~(\ref{r02}) and (\ref{E02}).
}
\end{figure}


\subsubsection*{Confinement and tilt energies}

A related set of power laws govern the change in the average optimal
score $\overline{S_0}$ if the alignment paths are subject to {\em constraints}.
For example, the constraint $-r_c/2 < r_0(t) < r_c/2$ artificially
confines the paths to a strip of width
$r_c$ on the alignment grid. This decreases the optimal score
$\overline{S_0}$ since the path $r_0(t)$
is cut off from random agglomerations of matches outside the strip. For long
sequences, this {\em confinement cost} becomes proportional to $N$, and
the average confinement cost per unit of $t$ is
\begin{equation}
\delta E_c(r_c; \gamma) \equiv \frac{\overline{S_0} (r_c; N, \gamma) -
                        \overline{S_0} (N,\gamma) }{N} < 0.
\end{equation}
It obeys the scaling law
\begin{equation}
\delta E_c (r_c; \gamma) \simeq - C(\gamma) \, r_c^{-1} \;,
 \label{Ec}
\end{equation}
with all the parameter dependence contained in the prefactor $C(\gamma)$.
This relation is valid in the asymptotic regime of strong confinement,
i.e., for sequences long enough that their unconstrained mean square
fluctuations
$\overline{ (\Delta r_0)^2(N)}$ exceed the scale $r_c^2$. According to
(\ref{r02}), this condition is satisfied for $N \gg r_c^{3/2} \,t_0 (\gamma)$.

In a similar way, the alignment may be constrained by restricting
both ends of the alignment path to given values of $r$, for example,
$r(0) = 0$ and $r(N) = r_0$. This forces an average tilt
$\theta = r_0/N$ upon the alignment path, thereby increasing
its number of gaps and decreasing its number of matches. The resulting
{\em tilt cost} is again proportional to $N$ for long sequences,
and the average tilt cost per unit of $t$,
\begin{equation}
\delta E_t (\theta; \gamma) \equiv
   \frac{ \overline{S_t} (\theta; N, \gamma)
         - \overline{S_t} (0;      N, \gamma)}{N} < 0
\end{equation}
is given by
\begin{equation}
\delta E_t (\theta; \gamma) \simeq - D(\gamma) \theta^2 \;.
\label{Et}
\end{equation}
This power law is valid for long sequences ($N \gg t_0(\gamma)$)
and small tilt angles ($\theta^2 < t_0^{-2} (\gamma)$).
The scaling form of the confinement and tilt energies has been
verified numerically, see Fig.~5.

\begin{figure}
\begin{center}
\epsfig{figure=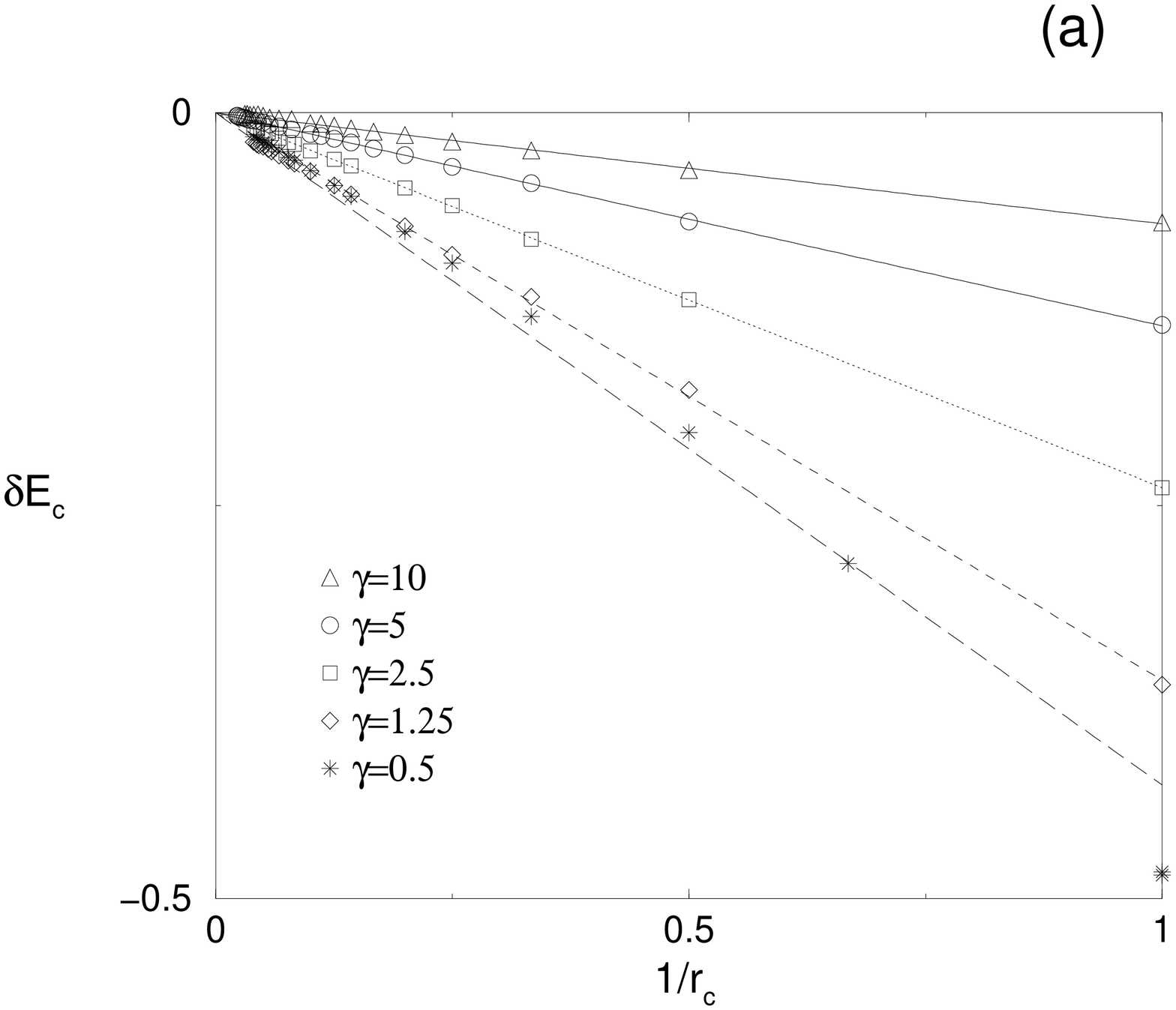,width=2.5in,angle=270}\hspace{0.5in}
\epsfig{figure=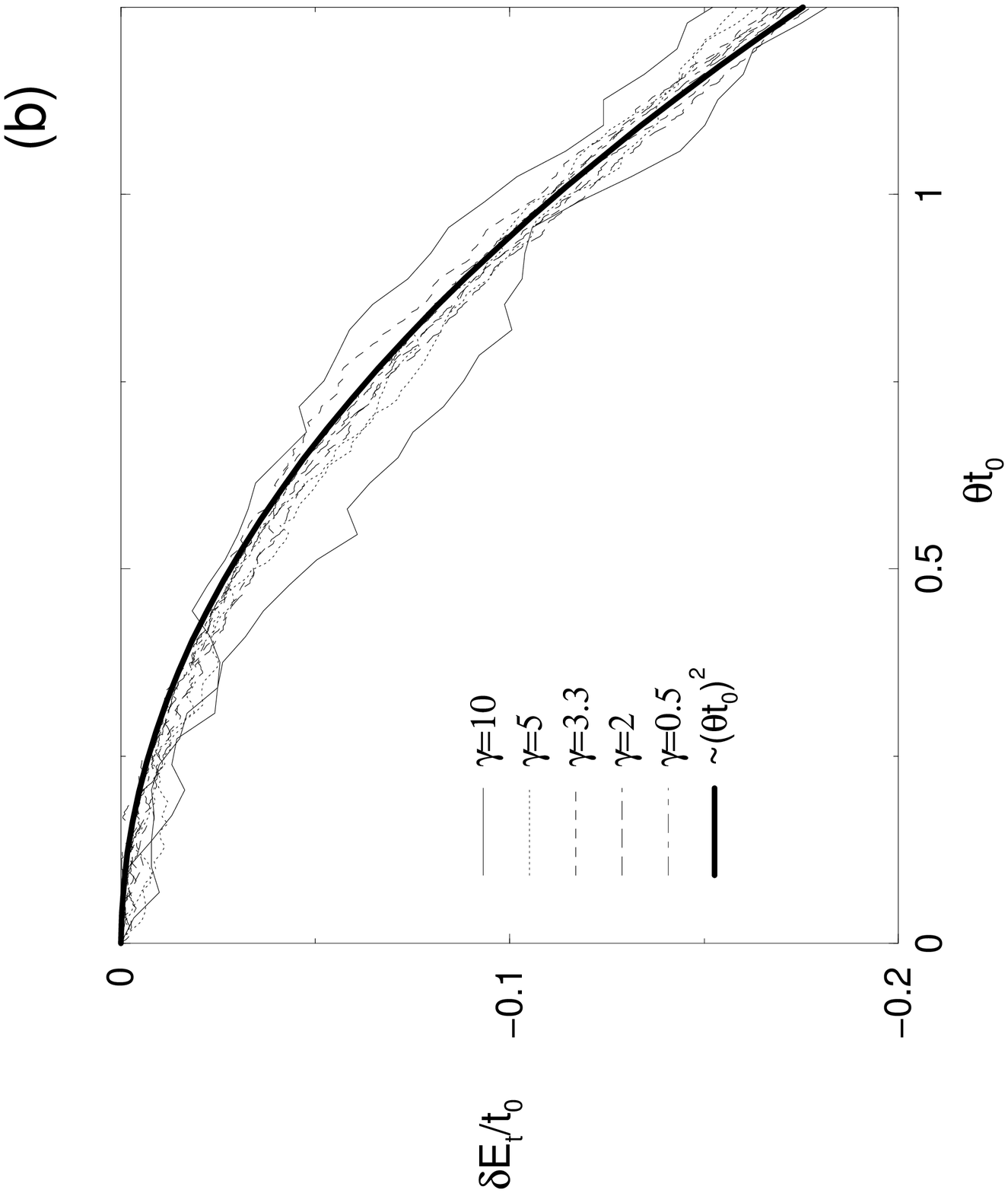,width=2.5in,angle=270}
\end{center}
\baselineskip=12pt {\small Fig.~5:
(a) Confinement cost $\delta E_c(r_c;\gamma)$ for optimal alignment paths
in a corridor $-r_c < r_0(t) < r_c$, taken from an ensemble of 200
mutually uncorrelated random sequences.
(b) Scaled tilt cost $\delta E_t (\theta; \gamma) / t_0(\gamma)$ as a
function of the scaled tilt $\theta \,t_0(\gamma)$,
for the same ensemble of sequences. The curves describe asymptotic
power laws with universal exponents and  and $\gamma$-dependent amplitudes,
as given by Eqs.~(\ref{Ec}) and (\ref{Et}).
}
\end{figure}

\begin{figure}
\centerline{\epsfig{file=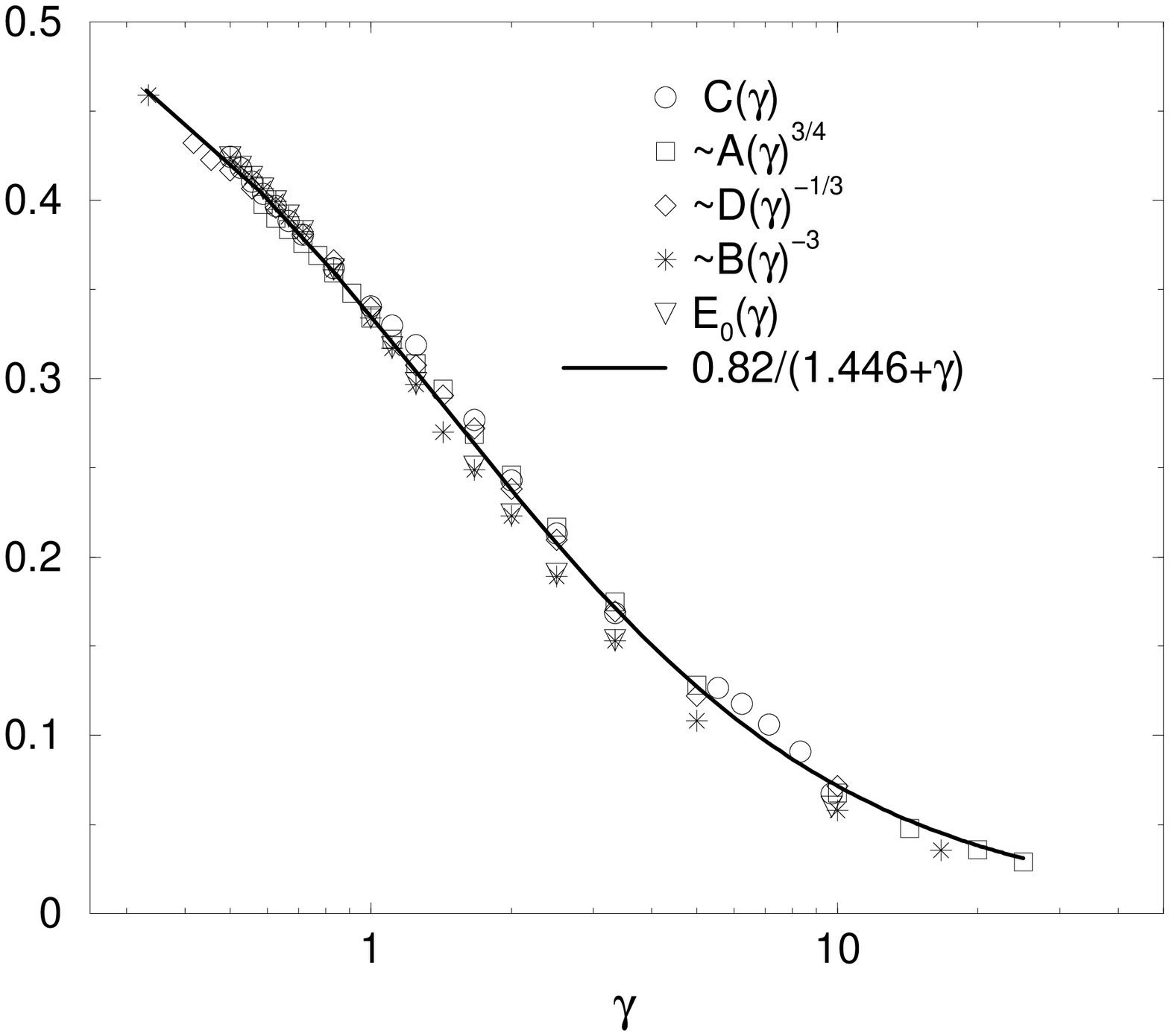,width=3.0in,angle=270}}
\baselineskip=12pt {\small Fig.~6:
Parameter dependence of the amplitudes
$A(\gamma)$, $B(\gamma)$, $C(\gamma)$m $D(\gamma)$, and $E_0(\gamma)$,
together with a fit curve of the form (\ref{Cg}). }
\end{figure}


\subsubsection*{Parameter dependence}

The scaling laws (\ref{r02}), (\ref{E02}), (\ref{Ec}) and (\ref{Et})
all have the same structure: they are power laws with universal
exponents and parameter-dependent
amplitudes. The scaling theory predicts not only the values of
the exponents but also universal relations between the amplitudes.
We have
\begin{equation}
A^{3/4} (\gamma) \propto
B^{-3} (\gamma) \propto
C (\gamma) \propto
D^{-1/3} (\gamma) \;,
\label{ampl}
\end{equation}
the tildes indicating proportionality factors independent of $\gamma$
and of order 1.
The amplitudes are monotonic functions of $\gamma$, which become
independent of $\gamma$ for $\gamma < \gamma _0$. Their asymptotic
behavior for large $\gamma$ can be calculated (Hwa and  L\"{a}ssig, 1996),
resulting in $C(\gamma) \sim \gamma^{-1}$. Indeed, we find this amplitude
to be well approximated by the form
\begin{equation}
C(\gamma) \sim (\gamma + \mbox{const.})^{-1}
\label{Cg}
\end{equation}
in the entire interval $\gamma > \gamma_0$. Our numerical data
verifying Eqs.~(\ref{ampl}) and (\ref{Cg}) are shown in Fig.~6.
We also show numerical results for the the average score per unit of $t$, i.e.,
the function $E_0(\gamma)$ in Eq.~(\ref{ave}). We find
$E_0(\gamma) \sim C(\gamma)$ approximately.

\section{Alignment of Correlated Sequences}

\subsubsection*{Displacement fluctuations of the evolution path}

As discussed in Section 2, the mutual correlations between sequences
are encoded in their evolution path, which is represented by the
evolution path $R(t)$ on the alignment grid. This path has displacement
fluctuations due to the random distribution of insertions and deletions,
see Figs.~2(c) and~3. However, the statistics
of these fluctuations is different from that of the alignment paths
discussed in the previous Section. Since the evolution is modeled as
a Markov process, the displacement
$\Delta R(t_1 - t_2) \equiv R(t_1) - R(t_2)$ has the mean square
\begin{equation}
\overline{\Delta R(t)^2} = q  |t|
\label{R2}
\end{equation}
characteristic of a Markov random walk, with
$q$ given by Eq.~(\ref{q}). The overbar denotes
an ensemble average over realizations of the evolution process
with given values of $U$ and $q$. The ensemble average
(\ref{R2}) equals the auto-correlation function of
a single sufficiently long evolution path $R(t)$ defined in analogy
to (\ref{auto}).

We may compare the fluctuations $\overline{\Delta R(t)^2}$
 of the evolution
path for {\em correlated} sequences with the
fluctuations $\overline{\Delta r_0(t)^2}$ of the optimal alignment path for
{\em uncorrelated} sequences (Drasdo et. al., 1997). 
This defines a  scale $\tilde t$,
where these fluctuations are of the same order of magnitude:
$\overline{(\Delta R(\tilde t))^2} =
\overline{(\Delta r_0(\tilde t))^2}
\equiv \tilde r^2$.
{From} Eqs.~(\ref{r02}) and (\ref{R2}), we obtain
\begin{equation}
\tilde t (\gamma, q) =  q^3 / A^6 (\gamma) \;, \hspace{1cm}
\tilde r (\gamma, q) =  q^2 / A^3 (\gamma) \;.
\label{rtilde}
\end{equation}
We call the scales (\ref{rtilde}) {\em roughness
matching} scales.
For $|t| < \tilde t (\gamma, q)$, the displacement of
the evolution path exceeds that of the optimal alignment path,
while for $|t| > \tilde t (\gamma, q)$, the
displacement of the alignment path becomes dominant.


\subsubsection*{Scaling theory for correlated sequences}

For sequences with mutual correlations (i.e., $U > 0$), the morphology
of the optimal alignment path $r_0(t)$ and the score statistics are more
complicated than for uncorrelated sequences since in addition to the
random matches, there are now the native matches along the evolution
path $R(t)$. Due to these competing score contributions, the problem
seems to be beyond the means of any rigorous mathematical approach.
However, it turns out that the statistics of {\em weakly} correlated sequences
is described with remarkable accuracy by the scaling theory developed in
the previous Section.

Consider a pair of weakly correlated sequences of length $N \gg 1$
 with an optimal alignment
of finite fidelity $\cF > 0$ at a given value of $\gamma$.
Since the optimal alignment path $r_0(t)$ and the evolution path $R(t)$
have a finite fraction of common bonds, the displacement fluctuations
of $r_0(t)$ remain confined to a ``corridor'' centered around the path
$R(t)$ (see Fig.~2(c)). The width $r_c$ of this corridor can be defined by
the mean square {\em relative} displacement
\begin{equation}
r_c^2 \equiv \overline{ (r_0(t) - R(t))^2} \;,
\label{rc2}
\end{equation}
which can again be understood as an ensemble average or equivalently
as an average over $t$ for a single pair of long sequences. To see this
equivalence, we note that by Eq.~(\ref{r02}), the width $r_c$ defines
a corresponding scale in $t$ direction, $t_c = r_c^{3/2} \, t_0 (\gamma)$.
One can show that $t_c$ is a {\em correlation length}; i.e., points on
the alignment path with $|t_2 - t_1| > t_c$ are essentially uncorrelated.
Averaging over uncorrelated regions of the alignment path generates
the ensemble underlying Eq.~(\ref{rc2}) even for a single pair of
sequences if they are sufficiently long, i.e.,  $N,N' \gg t_c$.

By confining the alignment path $r_0(t)$ to a corridor,
mutual correlations act as a  constraint on its displacement fluctuations.
This leads to a score cost as discussed in the Section~3.
However, the constraint
cost must be outweighed by the  gain in score due to the native matches,
resulting in a net score gain per unit of $t$,
\begin{equation}
\delta E(U, q, \gamma)
 \equiv \frac{\overline{S(N,U,q,\gamma)} -
              \overline{S_0(N,\gamma)      }
             }{N} > 0 \;,
\label{ionE}
\end{equation}
with $\overline{S_0(N,\gamma)}$ denoting the average score of uncorrelated
sequences.

We now calculate the confinement length $r_c(U,q,\gamma)$ and
the score gain $\delta E(U,q, \gamma)$ in a variational approach,
treating $r_c$ as an {\em independent} variable to be determined
a posteriori from an extremal condition. We stress that this approach
is not exact; the main approximation consists in treating $r$ and $t$
as continuous variables.

The constraint cost per unit of $t$ imposed by the evolution path $R(t)$
involves terms of the form discussed in the previous Section:
(i) If the path $r_0(t)$ is confined to a corridor of width $r_c$ around
the fluctuating path $R(t)$, the tangent to $r_0(t)$ has a  typical
tilt $\theta \sim q / r_c$ with respect to the diagonal of the
alignment grid, implying a tilt cost
\begin{equation}
\delta E_t (r_c; q,\gamma) \sim - D(\gamma)
\left(\frac{ q}{r_c}\right)^2 \;.
\label{Et.corr}
\end{equation}
(ii) The confinement cost to an untilted corridor of width $r_c$ is
$\delta E_c = C/r_c$. The tilt reduces the effective width of the
corridor so that the confinement cost takes the form
\begin{equation}
\delta E_c(r_c; q,\gamma) \sim - C(\gamma)
\frac{1 + q / [C^2(\gamma) r_c]}{r_c}
\;.
\label{Ec.corr}
\end{equation}
On the other hand, the gain in score per unit of $t$ due to the native
matches is simply $\delta E_n = U \cF$, as it is clear from
 the definition of the
fidelity ${\cal F}$. We need to express $\cF$ in terms
of $r_c$. Naively one would expect $\cF \sim 1/r_c$. A detailed analysis
shows that this is correct up to a logarithmic correction (Hwa and  Nattermann,
1995, Kinzelbach and  L\"{a}ssig, 1995, Hwa and  L\"{a}ssig, 1996)
leading to
\begin{equation}
\delta E_n (r_c; U) \sim U \, \frac{1 + \log r_c}{r_c} \;.
\label{En}
\end{equation}
The net score gain is the sum of these contributions,
$\delta E = \delta E_c + \delta E_t + \delta E_n$.
The resulting equation can be simplified by using the scaled variables
$x = C/U$,
$y = q/U^2$, and
$\delta {\cal E} = \delta E/U$. Absorbing all unknown proportionality
factors into their definition, we obtain the scaled energy gain
\begin{equation}
\delta {\cal E} (r_c; x,y) =
  - \frac{x}{r_c} -
  \frac{y}{x} \left (1 + \frac{y}{x^2} \right ) \frac{1}{r_c^2} +
  \frac{1 + \log r_c}{r_c} \;.
\label{Exy}
\end{equation}
Maximizing (\ref{Exy}) then determines
the actual value of $r_c (x,y) = r_c (U,q,\gamma)$:
\begin{equation}
\delta {\cal E}(x,y) = \max_{r_c} \delta {\cal E} (r_c; x,y) \;.
\label{rcmax}
\end{equation}

Fig.~7 shows numerical data for the fidelity
$\cF(x,y) = \cF(r_c(x,y); x,y)$ and score gain
$\delta {\cal E}(x,y)$ obtained from single sequence pairs with
various values of $U,q$ and $\gamma$. As expected from this scaling theory,
the data points for different parameter sets $(U,q,\gamma)$ corresponding
to the same $(x,y)$ collapse approximately.
This data collapse will be useful for similarity detection.

\begin{figure}
\begin{center}
\epsfig{figure=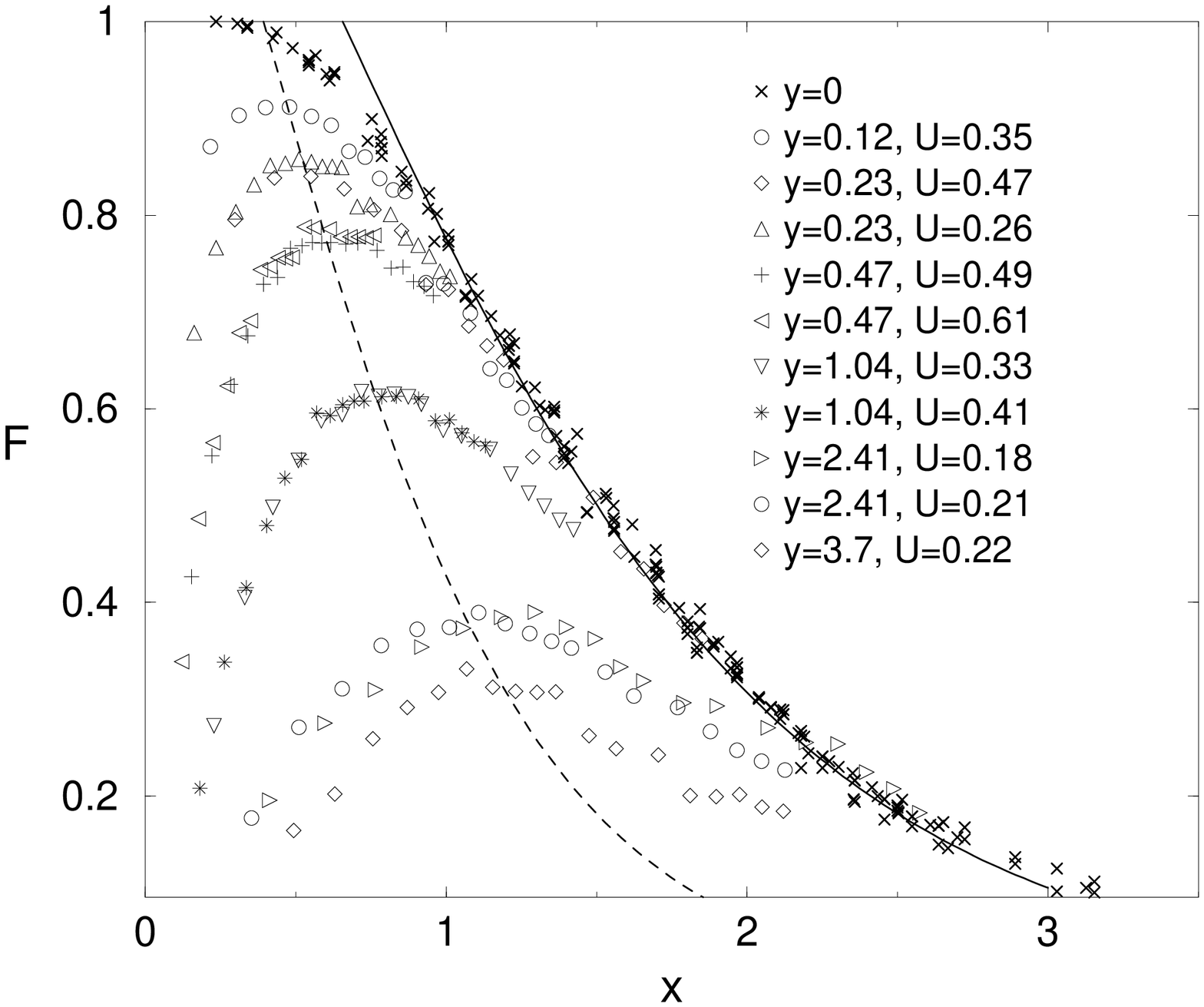,width=3.5in,angle=270} \\
\hspace*{-0.09in}\epsfig{figure=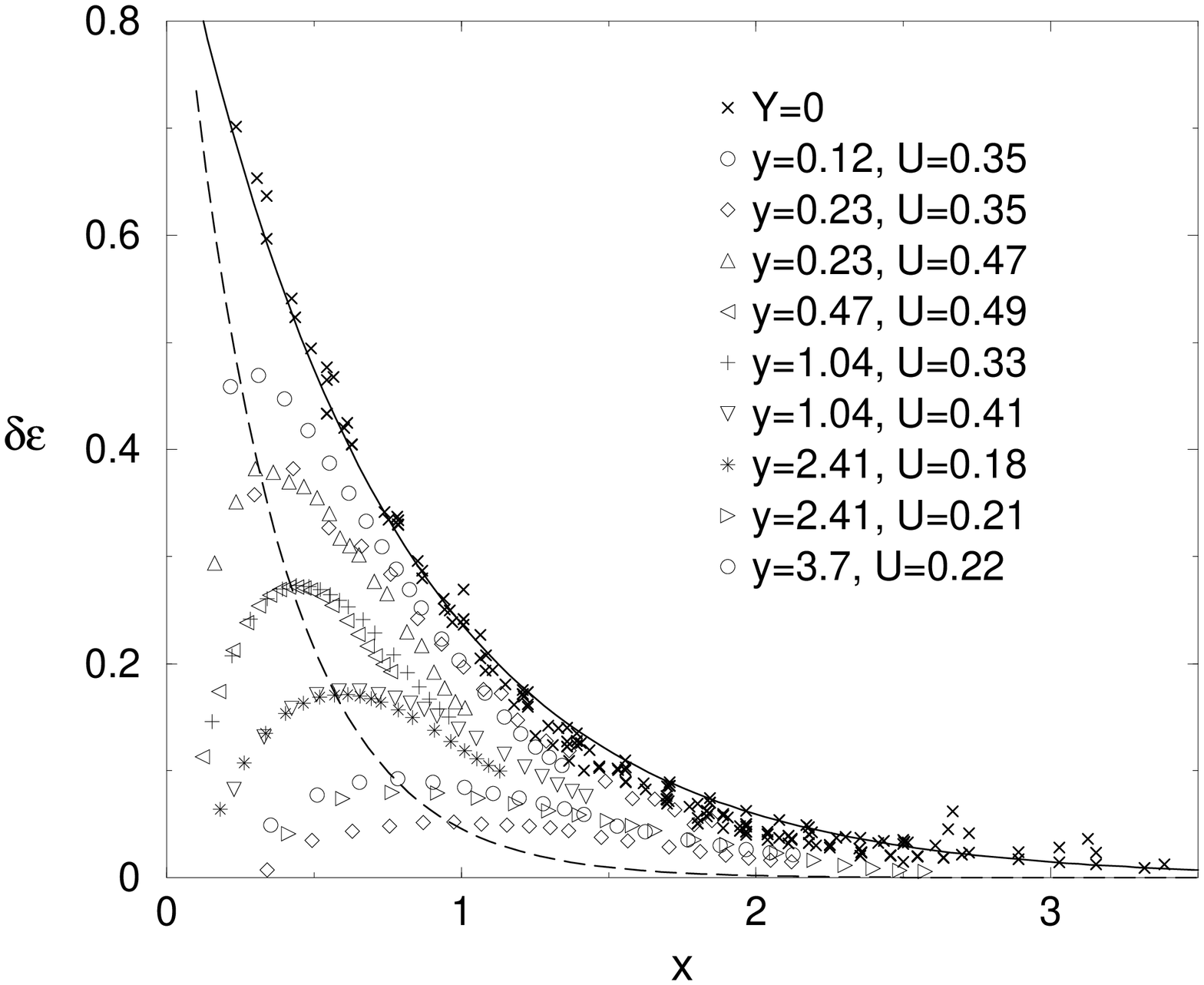,width=3.48in,angle=270}
\end{center}
\baselineskip=12pt {\small Fig.~7:
(a) Fidelity $\cF(x,y)$ and
(b) score gain $\dE(x,y)$ obtained from single sequence pairs with
various evolution parameters $U, q$ and alignment parameters $\gamma$.
The data for different $(U,q,\gamma)$ corresponding to the same
$(x,y)$ collapse approximately, as predicted by the scaling theory.
The lines are the theoretical loci of the
maxima $(x^*(y), \cF^*(y))$ (short-dashed),
$(x^s(y), \dE^s(y))$ (long-dashed) and
the theoretical limit curves $\cF(x, 0)$, $\dE(x,0)$ (solid).
}
\end{figure}

\subsubsection*{Alignment parameter optimization}

The numerical fidelity and score patterns of Fig.~7 have clear maxima
$\cF^*(y) \equiv \cF(x^*(y),y)$ and
$\dE^s(y) \equiv \dE(x^s(y),y)$, attained at points
$x^*(y)$ and $x^s(y)$. Fig.~7 also shows the loci of these
maxima, $(x^*(y), \cF^*(y))$ and $(x^s (y), \dE^s (y))$, as well as
the limit curves $\cF^*(x,0)$ and $\dE^s (x,0)$ obtained
from the scaling theory (i.e., from Eqs.~(\ref{Exy}) and (\ref{rcmax})
solved numerically). The theory is seen to predict the functional
form of these curves in a reasonable way,
 except in the region $\cF \sim 1$ (i.e.,
$r_c \sim 1$) where the continuum approximation valid in the
regime of  {\em weak} similarity breaks down.
(The unknown $\gamma$-independent proportionality
factors for the scaling variables $x$, $y$, $\dE$ and for $\cF$
have been determined by fits to the data.)

The functions $x^*(y)$, $x^s(y)$, $\cF^*(y)$, and $\dE^s(y)$ shown in Figs.~8
(a) and (b) encode in an efficient way the dependence of the
fidelity and score maxima on the alignment parameter {\em and} on
the evolution parameters.
%
Furthermore, it follows from the (numerical) solution of (\ref{Exy}) and
(\ref{rcmax}) that the confinement length $r_c^*(y) \equiv r_c(x^*(y),y)$
at the point of maximal fidelity satisfies the approximate relation
\begin{equation}
r_c^* (y)  \sim  \tilde r (x^*(y),y)
\end{equation}
in the biologically interesting regime of small $q$ and moderate $U$
($0 < y < 5$).
The optimal confinement length is thus proportional
to the roughness matching scale (\ref{rtilde}) at that point.
Hence, this scaling theory is in accordance with the qualitative picture
of Section 2: At $x^*(y)$, the fluctuations of the optimal alignment
path $r_0(t)$  just match those of the evolution path $R(t)$ (see Fig.~2(b)).
The shortcut regime (Fig.~2(c)) corresponds to the ascending branch
($x < x^*(y)$) of the fidelity curves in Fig.~7(a), while the
random fluctuation regime (Fig.~2(a)) corresponds to the descending
branch ($x > x^*(y)$).

\begin{figure}
\begin{center}
\epsfig{figure=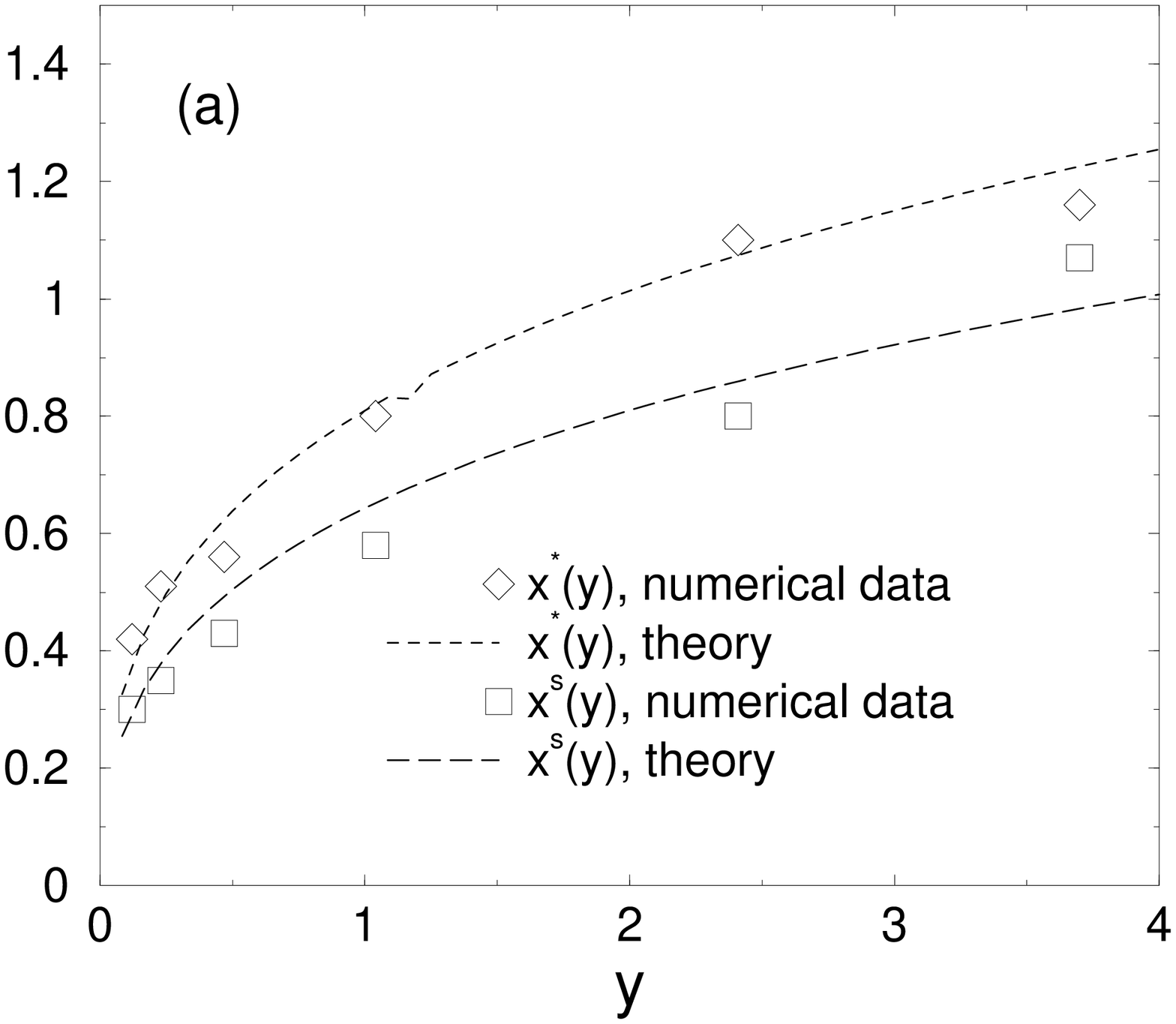,width=2.5in,angle=270}
\epsfig{figure=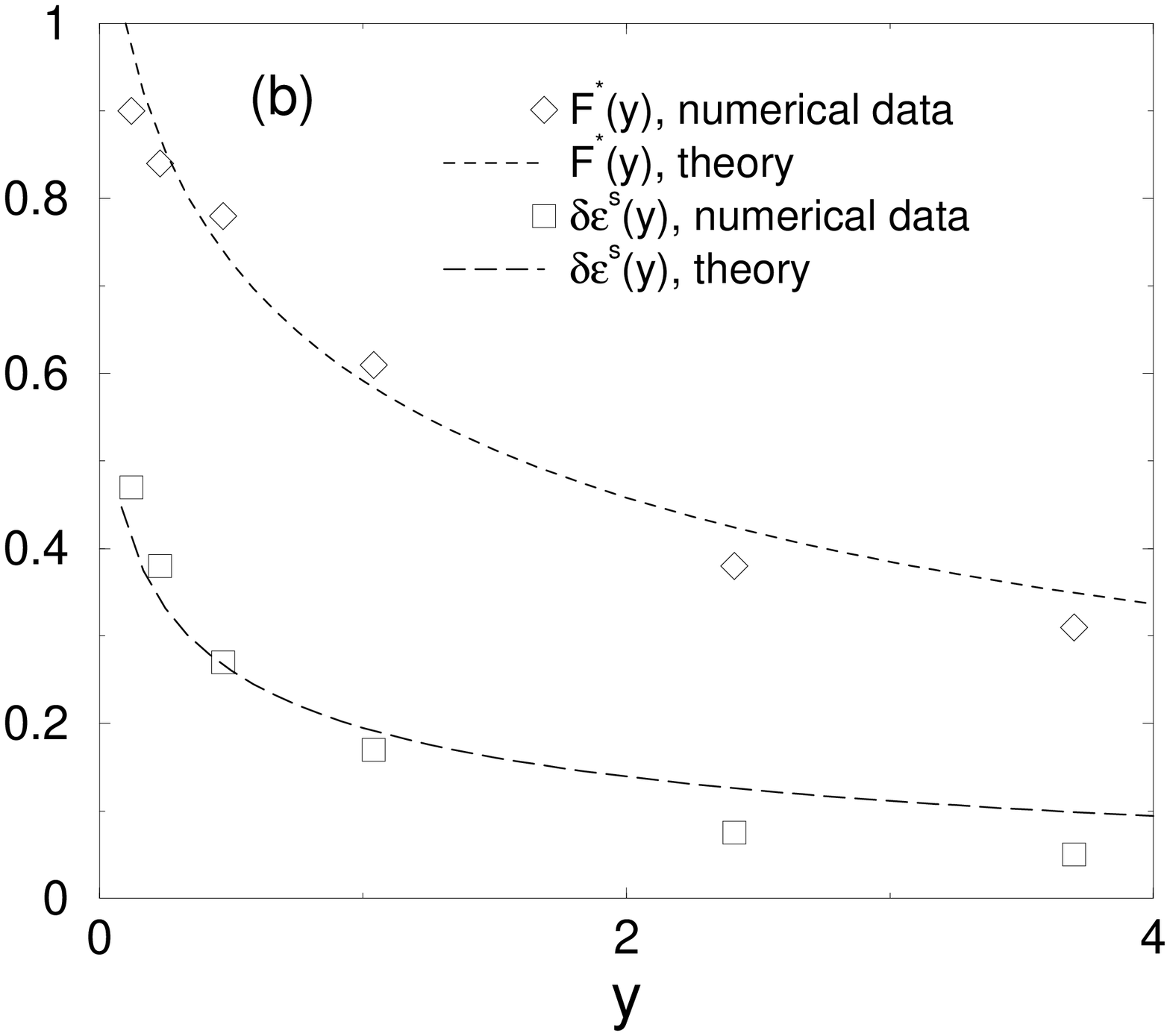,width=2.5in,angle=270}
\end{center}
\baselineskip=12pt {\small Fig.~8:
Alignments of maximal fidelity and of maximal score gain.
Theoretical predictions for the curves
(a) $x^*(y)$, $x^s(y)$ and
(b) $\cF^*(y)$, $\dE^s(y)$, compared to numerical data obtained from fits
to the curves of Fig.~7. }
\end{figure}


\subsubsection*{Similarity detection}

The evolution process used in this paper
is closely related to a more realistic process for
the divergent evolution of {\em two} daughter sequences $Q^{(1)}$
and $Q^{(2)}$ from a closest common ancestor sequence $Q$.
Modeling the two evolution paths as independent
Markov processes with respective parameters $U_1,q_1$ and $U_2,q_2$,
one can show that the evolution path linking $Q^{(1)}$ and
$Q^{(2)}$ is again a Markov process with parameters $U = U_1 U_2$
and $q = q_1 + q_2 + O(q^2)$.

For practical alignments, however, the evolutionary parameters $U$
and $q$ are unknown. Since they enter the definition
of the basic variables $x$ and $y$, knowledge of the optimal
parameters $x^*(y)$ and $x^s(y)$
seems to be of little use for applications.
However, these parameters can be reconstructed from alignment
data, as we will now show for a specific example.

Consider three sequences $Q^{(1)}$, $Q^{(2)}$ and
$Q^{(3)}$ related by the evolution tree of Fig.~9(a).
The evolutionary distances $\tau_i$ are
defined in terms of the mutual similarity coefficients $U_{ij}$ by
\begin{equation}
- \log U_{ij} = \tau_i + \tau_j \hspace{1cm} (i,j = 1,2,3) \;.
\label{tau}
\end{equation}
We whish to determine $\tau_1, \tau_2$ and $\tau_3$
from pairwise alignments of the sequences\footnote{
In this example, we use effective indel rates
$ -\log (1 - q_{ij}) = \Gamma (\tau_i + \tau_j)$ with
$\Gamma = 0.2$, but this choice is not crucial.}.
Fig.~9(b) shows the
alignment data $\delta E_{ij}$ as defined in Eq.~(\ref{ionE})
for each of these pairs, plotted as a function of $C(\gamma)$.
To fit the data curve $\delta E_{ij} (C)$ to the corresponding
scaled score gain curve $\dE_{ij} (x)$ of Fig.~7(b), we have to divide
both axes of the diagram by $U_{ij}$. In this way, we can
determine the {\em a priori} unknown factors $U_{ij}$, and hence the
evolutionary distances $\tau_i$,  see Fig.~ 9(b).
For this example, we obtain
$U_{12} \approx 0.54$,
$U_{13} \approx 0.43$,
$U_{23} \approx 0.415$, and
$\tau_1 \approx 0.22$,
$\tau_2 \approx 0.33$,
$\tau_3 \approx 0.55$,
which is to be compared with the actual values
$\tau_1 = 0.27$, $\tau_2 = 0.38$, and $\tau_3 = 0.61$
used to produce the sequences.

\begin{figure}[t]
\begin{center}
\epsfig{figure=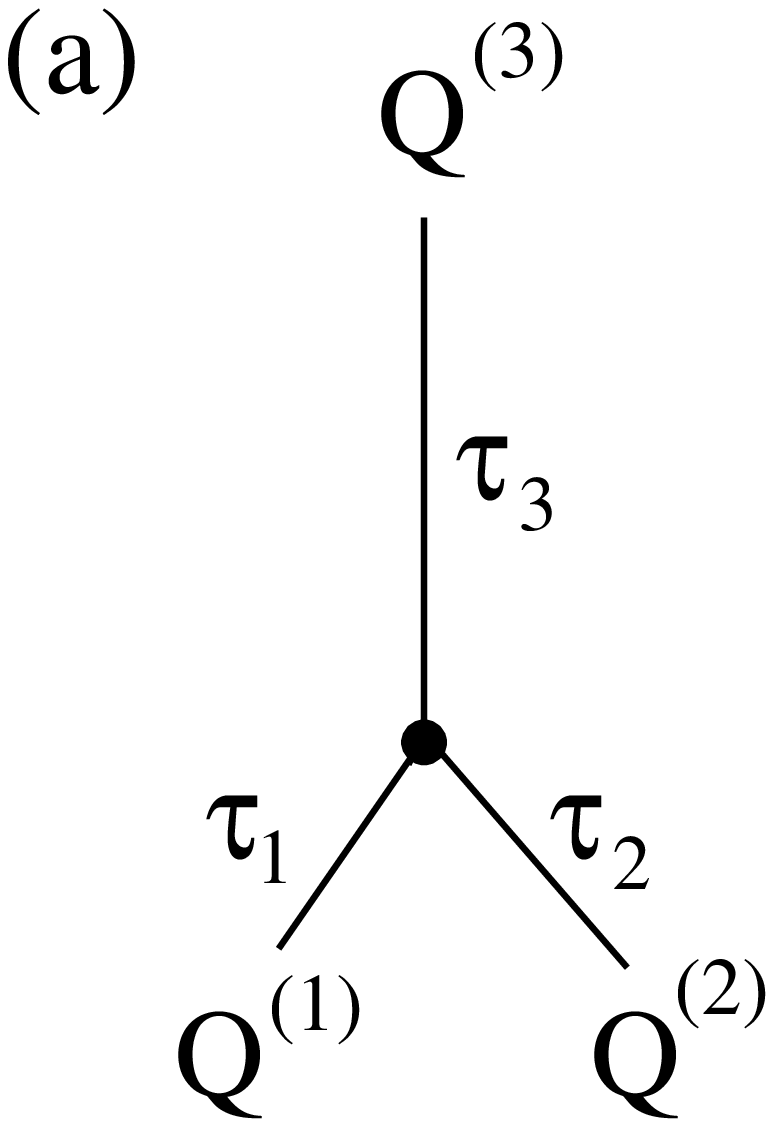,height=2.in,angle=0}\hspace{0.5in}
\epsfig{figure=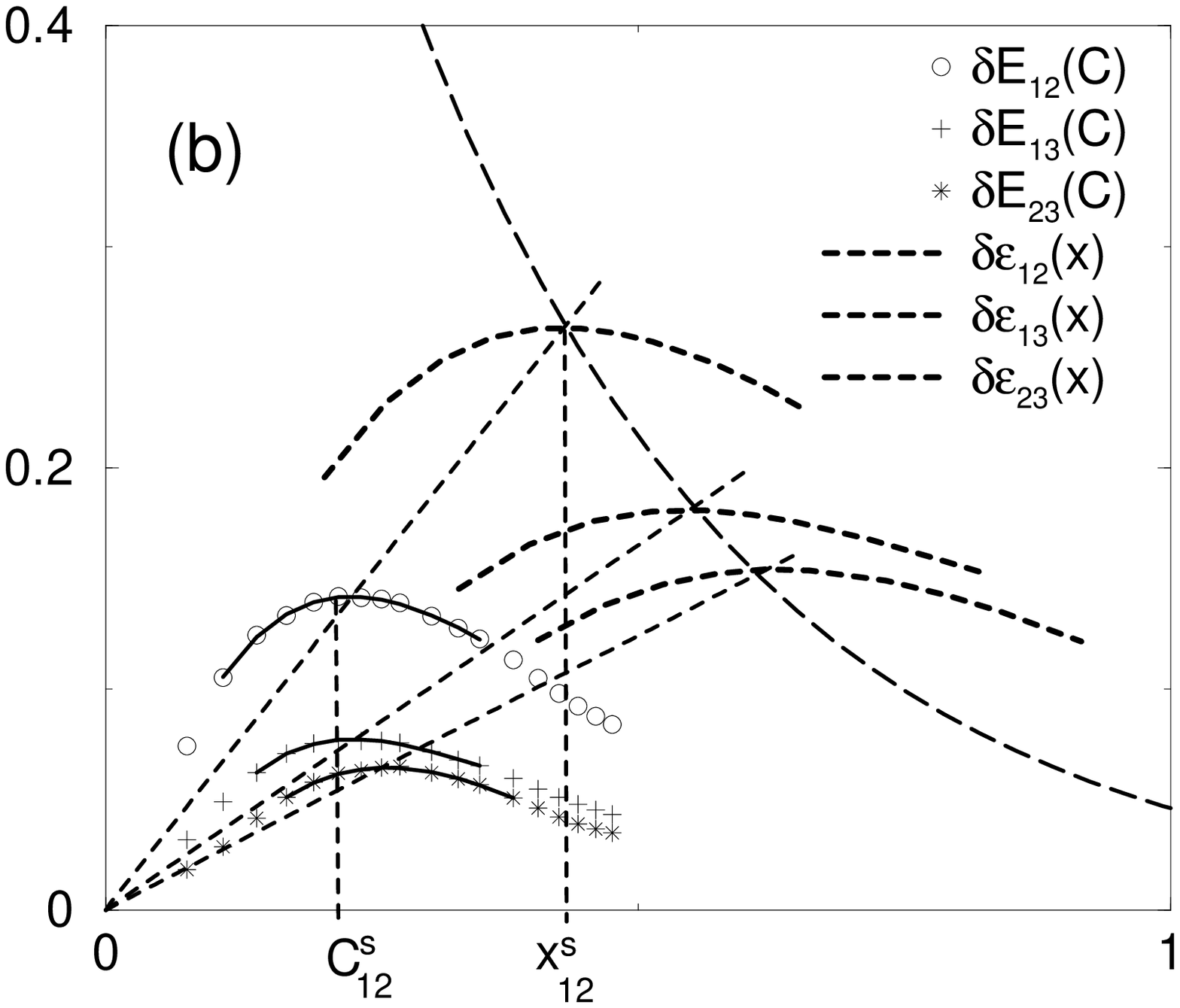,height=2.in,angle=0}
\end{center}
\baselineskip=12pt {\small Fig.~9:
(a) Evolution tree linking three sequences $Q^{(1)}$, $Q^{(2)}$,
and $Q^{(3)}$.
The sequences have evolutionary distances
$\tau_1$,
$\tau_2$, and
$\tau_3$ to the branching point of the tree, as defined by
Eq.~(\ref{tau}), and have lengths $N_1 \approx N_2 \approx N_3 \approx 5000$.
(b) Alignment data $\delta E_{12}$, $\delta E_{13}$ and $\delta E_{23}$
for pairwise alignments of the
sequences at different values of $\gamma$, shown as a function of $C(\gamma)$.
$\dE_{12}$, $\dE_{13}$, and $\dE_{23}$
obtained by rescaling the raw alignment data
by respective factors $U_{12}$, $U_{13}$, and $U_{23}$ such that
the maxima of the rescaled curves fall on the
theoretical locus $(x^s(y),\dE^s(y))$ (long-dashed  curve, cf.~Fig. 7(b)).
This determines the {\it a priori} unknown similarity coefficients
$U_{ij}$, and hence the evolutionary distances $\tau_i$.
}
\end{figure}

Finally, we construct the pairwise alignments of highest fidelity.
{From} Fig.~8(a), they are seen to satisfy the approximate relation
$C^*_{ij} / C^s_{ij} = x^*_{ij} / x^s_{ij}
\approx 1.2 $ for $0.1 < y < 4$.
With the values
$C^s_{12} \approx 0.23$,
$C^s_{13} \approx 0.225$, and
$C^s_{23} \approx 0.254$ read off from Fig.~9(b) and using Eq.~(\ref{Cg})
with the constants of Fig.~6,
we obtain the optimal alignment parameters
$\gamma_{12}^* \approx 1.52$,
$\gamma_{13}^* \approx 1.59$,
$\gamma_{23}^* \approx 1.25$. The scaled
score maxima
$\dE^s_{12} \approx 0.26$,
$\dE^s_{13} \approx 0.18$,
$\dE^s_{23} \approx 0.15$
determine the expected fidelities
$\cF^*_{12} \approx 0.75$,
$\cF^*_{13} \approx 0.58$,
$\cF^*_{23} \approx 0.52$
as seen from Fig.~ 8(b).
They are in good agreement with the actual maxima
$\cF^*_{12} = 0.8$,
$\cF^*_{13} = 0.65$,
$\cF^*_{23} = 0.55$
computed by comparing directly to the evolutionary paths.


\section{Discussion}

We have presented a statistical scaling theory for
global gapped alignments. Alignments of mutually uncorrelated
sequences are found to be governed by a number of
{\em universal scaling laws}: ensemble averages such as
the mean square displacement of the alignment path or the
variance of the optimal score follow power laws whose exponents
do not depend on the scoring parameters. The parameter dependence
is contained entirely in the prefactors. This universality
is comparable to the diffusion law describing a large variety
of random walk processes on large scales, the only parameter dependence
being the value of the diffusion constant. In contrast to diffusive random
walks, however, we find optimal alignment paths to be strongly non-Markovian on
all length scales due to random agglomerations of matches and mismatches.
Hence, the exponents take nontrivial values.
The scaling laws also govern the displacement statistics of a single
alignment path $r(t)$ and the associated statistics of partial scores,
which makes these concepts applicable to individual alignment problems.

This scaling theory is also relevant for
the statistics of mutually correlated sequence pairs. Two
important quantities are the {\em score gain} over uncorrelated sequences
and the alignment {\em fidelity}. Both quantities strongly depend on
the evolutionary parameters linking the two sequences and on the
alignment parameters. For a
simple Markovian evolution model and for linear scoring
functions, we have obtained a quantitative description of
this parameter dependence. In particular, the alignment parameter
of maximal fidelity turns out to be closely related to the
parameter of maximal score gain, which makes it possible to
construct the alignment of maximal fidelity from a systematic
analysis of score data. Moreover, the underlying evolutionary
parameters (the mutual similarity $U$ and the effective indel
rate $q$) can also be inferred from this analysis.

It is important to understand inhowfar the results of this paper
carry over to more refined algorithms for the alignment
of realistic sequences. The universal scaling laws for uncorrelated
sequences should prove to be very robust under changes of the
scoring function (such as scoring matrices distinguishing between
transitions and transversions) as well as changes in the sequences
(the number of different letters and their frequencies). As corroborated
by preliminary numerical results, such changes reduce to a different
parameter dependence of the amplitude functions $A,B, C,$ and $D$.
In particular, we find the universal scaling laws to be preserved
for the alignment of bona fide uncorrelated cDNA sequences, which
also validates the Markov model for single sequences. While not
affecting the asymptotic universality, some scoring functions
(for example, systems with affine gap cost distinguishing between gap
initiation and gap extension) may introduce intermediate regimes
where the score and fidelity curves are modified. Nevertheless,
the fidelity and the score gain remain
key quantities of an alignment, and their optimal
values are closely related. This makes it possible  to construct
 optimal alignments
on the basis of a statistical analysis of score data. This
link and the underlying scaling theory are also crucial to
the analysis of local alignment algorithms, as we have shown
recently (Hwa and  L\"{a}ssig, 1998; Drasdo {\it et al.}, 1998).

\vspace{\baselineskip}
\noindent {\it Acknowledgments.}
The authors are grateful to Stephen Altschul, Steven Benner, Richard Durbin,
Charles Elkan, Walter Fitch, Jeff Thorne, Martin Vingron, and Michael Waterman
for conversations and suggestions. TH acknowledges the financial support of an
A.~P.~Sloan Research Fellowship, an Arnold and Mabel Beckman Foundation
Young Investigator Award, and the hospitality of the Max-Planck Institute
at Teltow where much of the work was carried out.

\vfill
\newpage

\appendix

\clearpage

\section*{Appendix A: Evolution model}

The Markov process governing the evolution of a daughter sequence
$Q'$ from an ancestor sequence $Q$ is specified by the
flux diagram of Fig.~10.

\vspace{12pt}
\begin{figure}[h]
\centerline{\epsfig{file=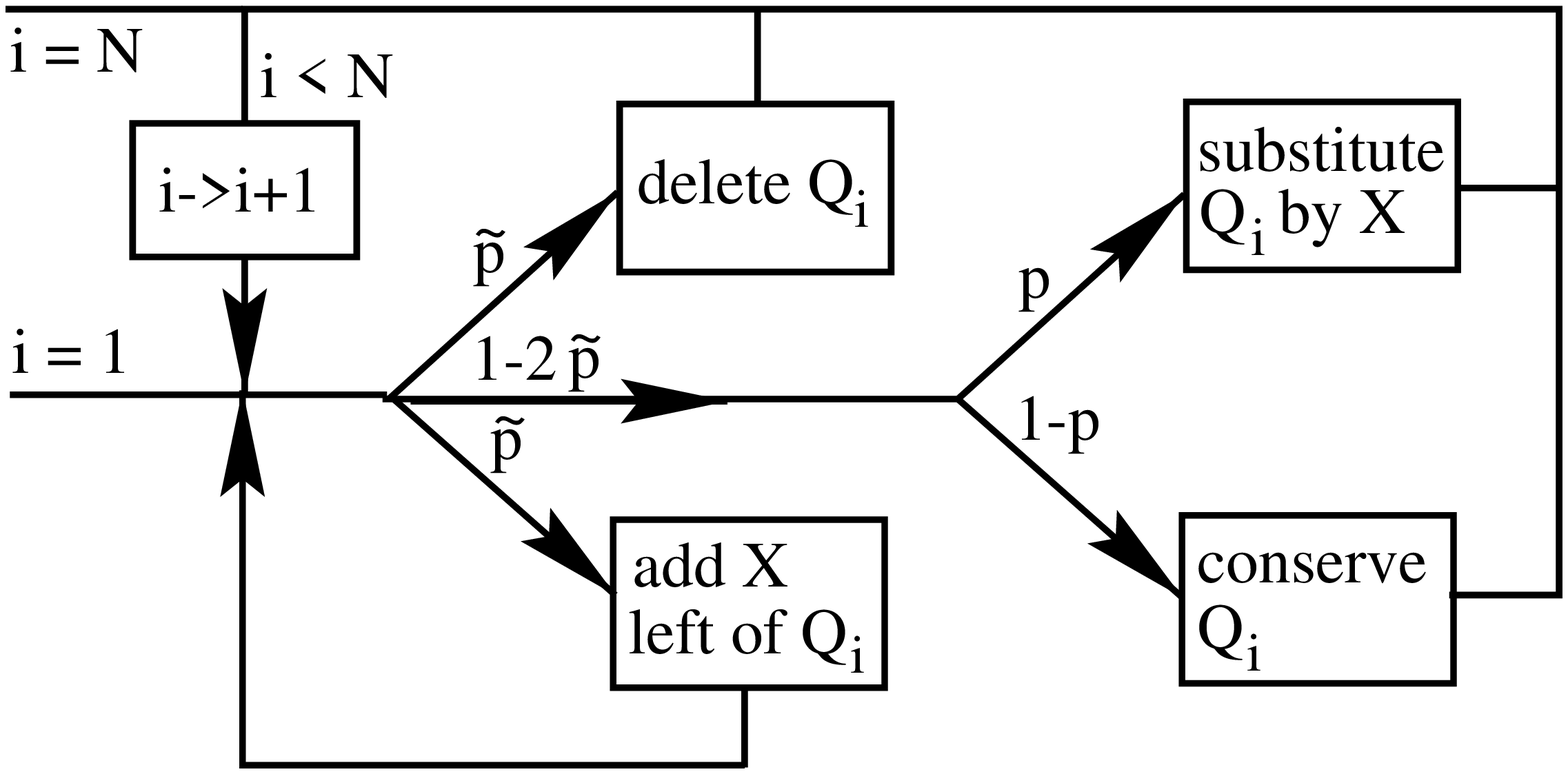,width=4in,angle=0}}
\baselineskip=12pt {\small Fig.~10:
Flux diagram of the Markov evolution process. A
realization generates a daughter sequence $Q' = \{Q'_j\}$ from
an ancestor sequence $Q = \{Q_i\}$.
The process is characterized by the insertion/deletion probability $\pt$
and the substitution probability $p$. $X$ denotes a random letter.
}
\end{figure}

The statistical properties of this Markov process are straightforward
to compute. Using the notation $t \equiv i + j$ and $R \equiv j - i$,
we find $R(t)$ to be a Gaussian random variable with
\begin{equation}
\overline{R(t)} = 0 \;, \hspace{1cm}
\overline{R^2 (t)} = q t \;,
\end{equation}
where $q$ is given by Eq.~(\ref{q}).
This implies in particular that the length $N'$ of the daughter
sequence is also a Gaussian random variable with
\begin{equation}
\overline{N'} = N \;, \hspace{1cm}
\overline{(N' - N)^2} = 2 q N \;.
\end{equation}

\section*{Appendix B: Alignment Algorithm}

The dynammic programming algorithm generates the score landscape
$S(r,t)$ for all grid points by the recursion relation
\begin{equation}
S(r,t) = \max \left \{ \begin{array}{l}
                       S(r-1, t-1) - \gamma \\
                       S(r+1, t-1) - \gamma \\
                       S(r,   t-2) + s(r,t)
                       \end{array} \right \}
\label{tm}
\end{equation}
with
\begin{equation}
s(r,t) = \left \{ \begin{array}{lll}
                  \sqrt{c - 1} \;\;\;& \mbox{ if } &
                  Q'_{(r+t)/2} = Q_{(r-t)/2} \\
                  - \frac{1}{\sqrt{c-1}} \;\;\;& \mbox{ if } &
                  Q'_{(r+t)/2} \neq Q_{(r-t)/2}
                  \end{array}  \right. \;.
\end{equation}
This recursion relation is evaluated in a restricted alignment grid
shown in Fig.~11, which limits the computing time  to
a value $\sim T \times L$. The value of $L$ is chosen according to the
specific application (see below). Along the strip, we use periodic boundary
conditions, i.e., $S(r-L/2,t) = S(r+L/2,t)$. (Similar results are obtained
for open boundary condition.)  Two types of initial
conditions are used depending on the specific application:
(i) $r(t \!=\!0) = 0$ (with $t \equiv i + j$) or
(ii) $S(t \! = \! 0) = 0$ (with $t \equiv i + j - L/2$). Evaluation of the
recursion relation stops at $t = T$. Hence, the optimal alignment
path $r_0(t)$ ends at the point $r_0 \equiv r_0(T)$ given by
$S(r_0,T) = \min_r S_0(r,T)$. If the score values $S_0(r,T)$ are degenerate
for different values of $r$, one of them is chosen at random. The
entire path $r_0(t)$ is then found by backtracking it from its endpoint
$r_0$. Degeneracies are again resolved by a random choices. This is
justified since degenerate optimal paths have a typical distance of
order $1$ only. For more precise formulations of this ``macroscopic
uniqueness'', see Fisher and Huse (1991), Hwa and Fisher (1994),
Kinzelbach and L\"{a}ssig (1995).

To compute the unconstrained fluctuations of optimal alignments for
uncorrelated sequences,
$L$ has to be sufficiently large so that the result becomes independent
of it: $L^2 \gg \overline{(\Delta r_0 (T))^2}$. The displacement
fluctuations $\overline{(\Delta r_0 (t))^2}$ and the tilt cost
$\delta E_t (\theta)$ are evaluated with the pinned initial condition~(i);
in the latter case, also the endpoint $r_0 = \theta T$ is pinned.
The score variance (\ref{Er2}) is computed with the initial condition~(ii).

On the other hand,
the confinement cost $\delta E_c (r_c)$ is determined by choosing
$L \equiv r_c$ and $T \gg L^{3/2} t_0(\gamma)$ so that the result
becomes independent of $T$ and of the initial condition.

For correlated sequences, we choose $L$ again large enough so that
the result becomes independent of it:
$L^2 \gg  \overline{(\Delta    R(T))^2} + r_c^2$.
For $T \gg t_c$, quantities
defined per unit of $t$ such as $\cF$ and $\delta E$ will also
become independent of the initial condition.

\begin{figure}
\centerline{\epsfig{figure=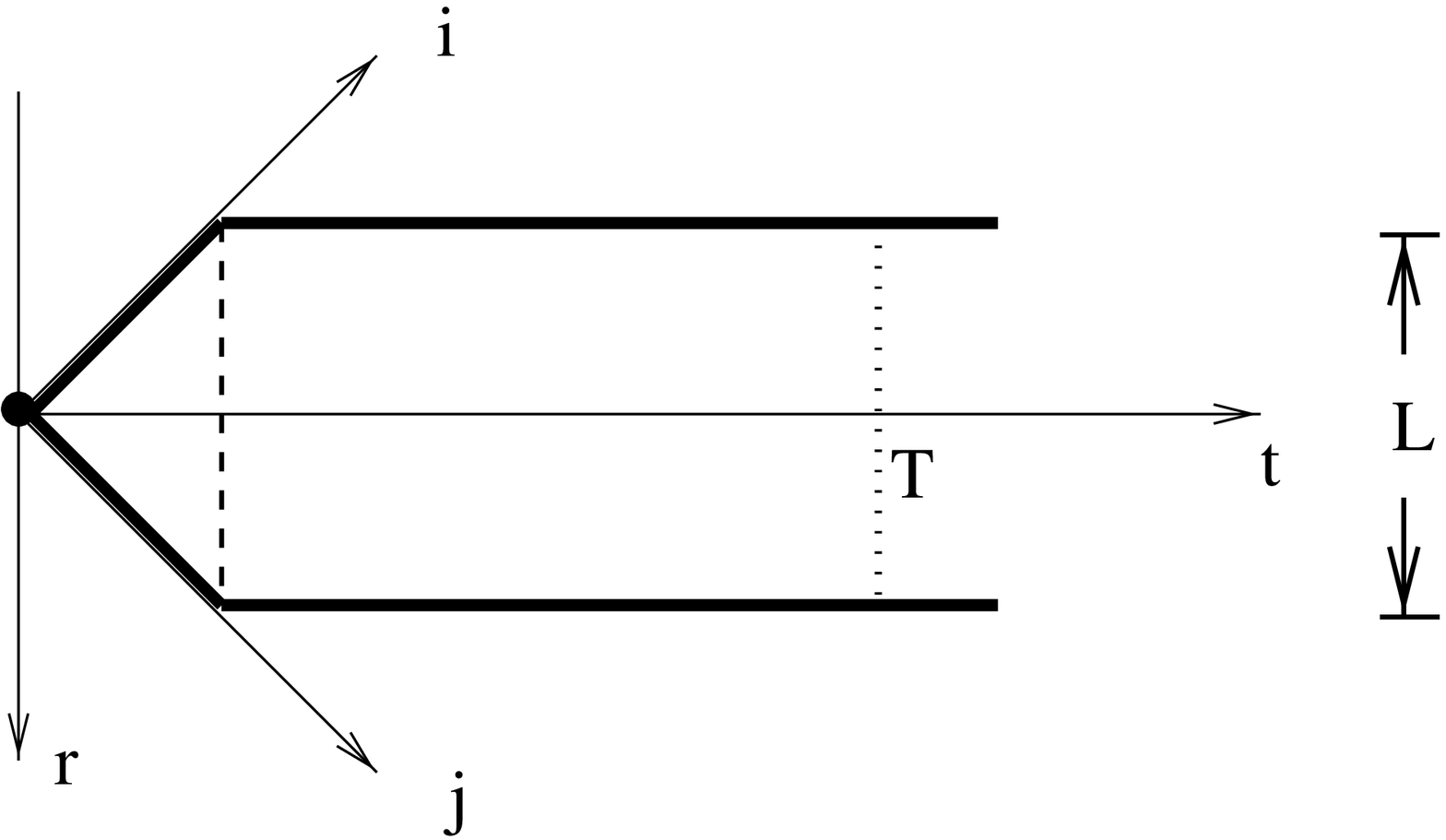,width=4in,angle=0}}
\baselineskip=12pt {\small Fig.~11:
Restricted alignment grid (bounded by thick lines)
used for the evaluation of the recursion relation (\ref{tm}).
With initial condition~(i), the alignment paths are pinned at
their initial point (dot) defined to be at $t = 0$. With initial
condition~(ii), the score is prescribed along the dashed line defined
to be at $t = 0$, namely $S(r, t \! = \! 0) = 0$.
}
\end{figure}


\newpage

\section*{References}

\noindent Altschul, S.F., Gish, W., Miller, W., Myers, E.W. and
Lipman, D.J. 1990. Basic local alignment search tool.
{\it J. Mol. Biol. 215 (3)}, 403 -- 10. \\

\noindent Altschul, S.F. 1993. A protein alignment scoring system
sensitive at all evolutionary distances. {\it J. Mol. Evol. 36 (3)}, 290 - 300.
\\

\noindent Arratia, R., Morris, P. and Waterman, M.S. 1988.
Stochastic scrabbles: a law of large numbers for sequence matching with scores.
 {\it J. Appl. Probab. 25} 106 -- 19. \\

\noindent Arratia, R. and Waterman, M.S., 1994.
A phase transition for the score in matching random sequences
allowing deletions.
{\it Ann of Appl. Prob. 4}, 200 -- 25. \\

\noindent Benner, S.A., Cohen, M.A. and Gonnet, G.H. 1993.
Empirical and structural models for insertions and
deletions in the divergent evolution of proteins.
{\it J. Mol. Biol. 229 (4)}, 1065 -- 82. \\

\noindent Bishop, M.J. and Thompson, E.A. 1986.
Maximum likelihood alignment of DNA sequences. {\it J. Mol. Biol. 190
(2)}, 159 - 65. \\

\noindent Cule, D. and Hwa, T. 1998.
Static and Dynamic Properties of Inhomogeneous Elastic Media on Disordered
Substrate.
{\it Phys. Rev. B.} in press. \\

\noindent Drasdo, D., Hwa, T. and L\"{a}ssig, M. 1997.
DNA sequence alignment and critical phenomena.
{\it Mat. Res. Soc. Symp. Proc.}  {\bf 263}, 75-80. \\

\noindent Drasdo, D., Hwa, T. and L\"{a}ssig, M. 1998.  A statistical theory of
sequence alignment with gaps,
submitted to {\it The Sixth International
Conference on Intelligent Systems for Molecular Biology}. \\

\noindent {From} MEDLINE; 96096722, cf Exposito J.Y., Boute N., Deleage
G., Garrone R.. 1995.  Characterization of two genes coding for a similar
four-cysteine
motif of the amino-terminal propeptide of a sea urchin fibrillar
collagen. Eur. J. Biochem. 234:59-65. \\

\noindent {From} MEDLINE; 94215495, cf. Fehon R.G., Dawson I.A.,
Artavanis-Tsakonas S. 1994.  A Drosophila homologue of membrane-skeleton
protein 4.1 is
associated with septate junctions and is encoded by the coracle
gene.
Development 120:545-557. \\

\noindent Fisher, D.S. and Huse, D.A. 1991.
Directed paths in a random potential.
{\it Phys.~Rev.~B 43 (13)}, 10728 - 10742. \\


\noindent Gusfield, D., Balasubramanian, K., and Naor, D.. 1992. {\it
Proceedings of the Third Annual ACM-SIAM Symposium on discrete Algorithms,
January 1992}, 432. \\

\noindent Hwa, T. and Fisher, D.S. 1994.
Anomalous fluctuations of directed polymers in random media.
{\it Phys. Rev. B 49}, 3136 -- 54. \\

\noindent  Hwa, T. and L\"{a}ssig, M. 1996.
Similarity detection and localization, {\it Phys. Rev. Lett. 76}, 2591 - 2595.
\\

\noindent  Hwa, T. and Nattermann, T. 1995.
Disordered induced depinning transition, {\it Phys.~Rev.~B 51},
455 - 469. \\

\noindent Hwa, T. and L\"{a}ssig, M.  Optimal detection of
sequence similarity by local alignment. Proc. of the Second Annual
Conference on Computational Molecular Biology (RECOMB98), in press.
E-print cond-mat/9712081. \\

\noindent  Karlin, S. and Altschul, S.F. 1990. Methods for assessing
the statistical significance of molecular sequence features by using general
scoring schemes. {\it Proc. Natn. Acad. Scie.  U.S.A. 87 (6)}, 2264 - 8. \\

\noindent  Karlin, S. and Altschul, S.F. 1993. Applications and
statistics for multiple high-scoreing segments in molecular sequences.
{\it Proc. Natn. Acad. Scie.  U.S.A. 90 (12)}, 5873 - 7. \\

\noindent
Koretke, K.K., Kutheyschulten, Z., Wolynes, P.G. 1996.
Self-consistently optimized statistical
mechanical energy functions for sequence structure alignment
{\it Prot. Sci. 5}, 1043-1059. \\

\noindent Kardar, M. 1987.
Replica Bethe ansatz studies of two-dimensional interfaces with quenched random
impurities. {\it Nucl. Phys. B 290}, 582 - 602. \\

\noindent Kinzelbach, H. and L\"{a}ssig, M. 1995. Depinning in a random
medium. {\it J.~Phys.~A: Math. Gen. 28}, 6535 - 6541. \\

\noindent Licea, C. and Newman, C.M. 1996.
Geodesics in two-dimensional first-passage percolation.
{\it Ann. Probab. 24}, 399 -- 410. \\

\noindent Licea, C., Newman, C.M., and Piza, M.S.T. 1996.
Superdiffusitivity in first-passage percolation.
{\it Probab. Theory Relat. Fields 106}, 559 -- 91. \\

\noindent
Onuchic, J.N., LutheySchulten, Z., Wolynes, P.G. 1997.
Protein folding funnels: the nature
of the transition state ensemble.
{\it Ann. Rev. Phys. Chem. 48}, 545-600, and references therein. \\

\noindent Needleman, S.B. and Wunsch, C.D. 1970.
A general method applicable to the search for similarities in
the amino acid sequence of two proteins.
{\it J. Mol. Biol. 48 (3)}, 443 - 53. \\

\noindent Pearson, W.R. 1991.
Searching protein sequence libraries: comparison of the sensitivity and
selectivity of the Smith-Waterman and FASTA algorithms.
{\it Genomics 11 (3)}, 635 - 650. \\

\noindent Smith, T.F. and Waterman, M.S. 1981.
Identification of common molecular subsequences.
{\it J. MOl. Biol. 147}, 195 -- 7. \\

\noindent Thorne, J.L., Kishino, H. and Felsenstein, J. 1991.
An evolutionary model for maximum likelihood
alignment of DNA sequence {\it J. Mol. Evol. 33 (2)}, 114 - 24 \\

\noindent Thorne, J.L., Kishino, H., and Felsenstein, J. 1992.
Inching toward reality: an improved likelihood model of sequence evolution.
{\it J. Mol.  Evol. 34 (1)}, 3 - 16. \\

\noindent Vingron, M. and Waterman, M.S. 1994. Sequence alignment and
penalty choice. Review of concepts, case studies and implications.
{\it J. Mol. Biol 235 (1)}, 1 - 12. \\
%

\noindent
Wang, J., Onuchic, J., Wolynes, P.G. 1996.
Statistics of kinetic pathways on biased rough energy landscapes
with application to protein folding. \\
{\it Phys. Rev. Lett. 76}, 4861-4864.

\noindent Waterman, M.S., Gordon, L. and Arratia, R. 1987.
Phase transitions in sequence matches and nucleic acid structure.
{\it Proc. Natl.  Acad. Sci. U.S.A. 84 (5)}, 1239 - 43. \\

%
\noindent Waterman, M.S. 1989. {\it In} Waterman, M.S, ed.,
{\it Mathematical Methods for DNA Sequences}.  CRC Press. \\

\noindent Waterman, M.S. Eggert, M. and Lander, E. 1992.
Parametric sequence comparisons.
{\it Proc. Natn.  Acad. Scie. U.S.A. 89 (13)}, 6090 - 3. \\

\noindent Waterman, M.S. 1994. {\it Introduction to Computational Biology},
Chapman \&  Hall. \\

\noindent Waterman, M.S. 1994.
Parametric and ensemble sequence alignment algorithms.
{\it Bull. Math. Biol. 56 (4)}, 743 - 767. \\

\noindent Zhang, M.Q. and Marr, T.G. 1995.
Alignment of molecular sequences seen as random path analysis.
{\it J. Theo. Biol. 174 (2)}, 119 - 29.
\end{document}